\documentclass[10pt,journal,compsoc]{IEEEtran}

\ifCLASSINFOpdf

\else

\fi
\usepackage{booktabs}
\usepackage{array}
\newcommand{\PreserveBackslash}[1]{\let\temp=\\#1\let\\=\temp}
\newcolumntype{C}[1]{>{\PreserveBackslash\centering}p{#1}}
\newcolumntype{R}[1]{>{\PreserveBackslash\raggedleft}p{#1}}
\newcolumntype{L}[1]{>{\PreserveBackslash\raggedright}p{#1}}
\usepackage{enumerate}
\usepackage{amsmath}
\usepackage{mathtools}
\usepackage{graphicx}
\DeclareGraphicsExtensions{.eps,.ps,.jpg,.bmp}

\usepackage{subfigure}
\usepackage{amssymb}
\usepackage{multirow}
\usepackage[noend]{algorithmic}
\usepackage{ifpdf}
\usepackage{verbatim}
\usepackage[numbers,sort&compress]{natbib}
\usepackage[ruled]{algorithm2e}

\usepackage{makecell,rotating,multirow,diagbox}

\begin{document}
	
	\title{Detect Professional Malicious User with Metric Learning in Recommender Systems}
	\author{Yuanbo~Xu,
		Yongjian~Yang,
		En~Wang$^\dag$,
		Fuzhen~Zhuang,
		Hui~Xiong,~\IEEEmembership{Fellow,~IEEE,}
		
		\IEEEcompsocitemizethanks{
			\IEEEcompsocthanksitem Y. Xu, Y. Yang, and $^\dag$E. Wang (corresponding author) are with the Department of Computer Science and Technology, Jilin University, Changchun, Jilin 130012, China. E-mail: {yuanbox, yyj, wangen}@jlu.edu.cn.
			\IEEEcompsocthanksitem F. Zhuang is with the Key Lab of Intelligent Information Processing of Chinese Academy of Sciences (CAS), Beijing 100190, China; and also with University of Chinese Academy of Sciences, Beijing 100049, China. E-mail: zhuangfuzhen@ict.ac.cn.
			\IEEEcompsocthanksitem H. Xiong is with the School of Business, Rutgers, the state university of New Jersey, USA. E-mail: hxiong@rutgers.edu.
			
	}}
	
	%
	%

	\markboth{Journal of \LaTeX\ Class Files,~Vol.~14, No.~8, August~2020}%
	{Shell \MakeLowercase{\textit{et al.}}: Bare Demo of IEEEtran.cls for IEEE Journals}
	%



	\IEEEtitleabstractindextext{%
		\begin{abstract}
			In e-commerce, online retailers are usually suffering from professional malicious users (PMUs), who utilize negative reviews and low ratings to their consumed products on purpose to threaten the retailers for illegal profits. PMUs are difficult to be detected because they utilize masking strategies to disguise themselves as normal users. Specifically, there are three challenges for PMU detection: 1) professional malicious users do not conduct any abnormal or illegal interactions (they never concurrently leave too many negative reviews and low ratings at the same time), and they conduct masking strategies to disguise themselves. Therefore, conventional outlier detection methods are confused by their masking strategies. 2) the PMU detection model should take both ratings and reviews into consideration, which makes PMU detection a multi-modal problem. 3) there are no datasets with labels for professional malicious users in public, which makes PMU detection an unsupervised learning problem. To this end, we propose an unsupervised multi-modal learning model: MMD, which employs Metric learning for professional Malicious users Detection with both ratings and reviews. MMD first utilizes a modified RNN to project the informational review into a sentiment score, which jointly considers the ratings and reviews. Then professional malicious user profiling (MUP) is proposed to catch the sentiment gap between sentiment scores and ratings. MUP filters the users and builds a candidate PMU set. We apply a metric learning-based clustering to learn a proper metric matrix for PMU detection. Finally, we can utilize this metric and labeled users to detect PMUs. Specifically, we apply the attention mechanism in metric learning to improve the model's performance. The extensive experiments in four datasets demonstrate that our proposed method can solve this unsupervised detection problem. Moreover, the performance of the state-of-the-art recommender models is enhanced by taking MMD as a preprocessing stage.
			
		\end{abstract}
		}
	
	\maketitle
	
	\section{Introduction}
	
	\IEEEPARstart{E-}{commerce} giants, such as Amazon, Jingdong, and Alibaba, have been thriving with the development of Internet technology, where millions of electronic retailers produce great wealth through selling commodities on the websites \cite{D4traver2008commerce}. For each day, billions of trades occur between retailers and consumers \cite{D5rutherford2016customer}. For the sake of improving the consumers' experience of online shopping, e-commerce websites usually allow consumers (we call them ``users'') to leave reviews and rank ratings on the commodities (we call them ``items"). To trade off the interests between retailers and users, e-commerce websites punish the retailers who receive a high percentage of negative reviews and low ratings from users\cite{D6akter2016big}. Being widely applied in almost all kinds of e-commerce websites, this feedback mechanism has been proved to be effective if all the users leave truthful and objective reviews or ratings.
	\begin{figure}[tbp]
		\centering
		\includegraphics[width=0.9\columnwidth]{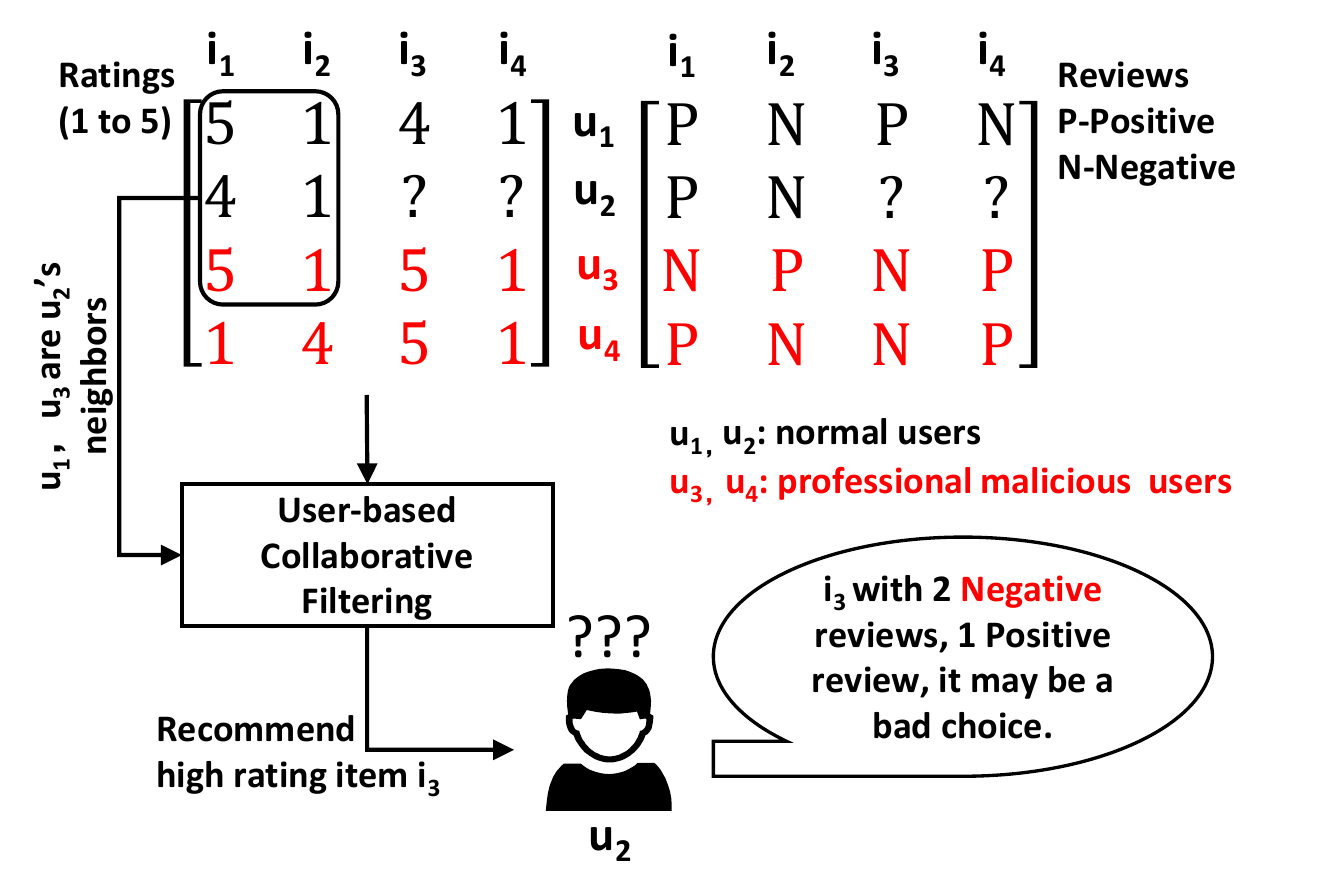}
		\caption{An example to indicate how professional malicious users confuse normal users and undermine the fairness of online e-commerce. In this example, $u_3$ and $u_4$ are professional malicious users, who give a high rating (5) for $ i_3 $, while also a negative review (N) for $ i_3 $. Hence, normal user $u_2$ is confused about recommendation $i_3$ by the fake negative feedbacks of professional malicious users $u_3$, $u_4$. More importantly, $u_3$'s distribution of ratings is the same as that of the normal user $u_1$, so traditional statistic outlier detections cannot detect this kind of professional malicious users for recommender systems.}
		\label{MMD1-1}
		\vspace{-15pt}
	\end{figure}
	
	However, in practice, there exist some malicious users (MU), who leverage this feedback mechanism to gain illegal profits \cite{D8si2018shilling,D9cai2019trustworthy}. For example, these malicious users first purposefully leave negative reviews and low ratings of their consumed products without any consideration of the commodities' quality. Then they blackmail the electronic retailers to make illegal profits; otherwise, they would leave more negative feedbacks, cheating e-commerce websites to punish the electronic retailers and confuse the normal users about the items in recommendations. As a result, these malicious users undermine the fairness of e-commerce. Moreover, their negative feedbacks will confuse the recommender systems (collaborative filtering-based models \cite{D14he2017neural} or content-based models \cite{D15su2016content}), leading to a chaotic recommendation for normal users, which is also named as shilling attacks \cite{D10zhou2016svm,D12xu2019detecting}.
	
	To address the above issues, e-commerce companies usually employ statistic outlier detection or shilling attack detection models \cite{D2kwak2017statistical,D3maronna2019robust,46tong2018shilling} to detect MUs, i.e., finding objective users who always give negative reviews or low ratings. However, there are some restrictions for these detection models: first, these models only tackle this problem from a methodological perspective and ignore the real-world scenarios. For example, most detection models ignore that there are some professional malicious users (PMUs), who can utilize masking strategies to avoid detection; second, they usually focus on filtering either fake ratings to improve recommendation models, or negative reviews for content-based models, which do not take both ratings and reviews into account. As a result, these models may be applied in limited application scenarios in recommender systems, but not proper for professional malicious user detections. 
	
	Different from malicious users, professional malicious users (PMU) typically adopt the following two masking strategies to avoid existing detections: 1) To avoid giving too many low ratings, they provide a high rating but a negative review. In this way, they can mislead the potential consumer who is browsing this review to decide whether to buy this item. 2) To avoid giving too many negative reviews, they provide a low rating but a positive review. In this way, they can explain to the outlier detection that their interactions are ``misoperations''. By applying the above two strategies, alternately, professional malicious users can disguise themselves as normal users. As shown in Fig.\ref{MMD1-1}, we give an example to indicate how professional malicious users confuse potential consumers and undermine the fairness of online e-commerce. In this paper, we focus on how to detect these PMUs with masking strategies in real-world scenarios by simultaneously analyzing their ratings and reviews.
	
	To detect PMUs, there are three significant challenges in recommender systems: 1) PMUs adopt masking strategies to act like normal users, which is difficult to be detected. 2) Detecting PMUs needs to analyze both ratings and reviews, which makes it a multi-modal problem. 3) Existing public datasets do not contain the PMU label, which makes this detection an unsupervised learning problem. To this end, we propose an unsupervised multi-modal learning model: \textbf{MMD}, which applies metric learning \cite{D21Wang2018Robust,D22Sui2018Convex,D23Zuo2017Distance} for professional malicious user detection with both ratings and reviews. The key to metric learning is utilizing different metrics (Euclidean distance or other metrics) to represent the relationships between entities \cite{D19Ye2018Fast,D20Li2018Semi}. MMD first utilizes Hierarchical Dual-Attention RNN (HDAN) \cite{D13xu2019neuo} to do user profiling with reviews and ratings. By catching the sentiment gap between reviews and ratings, we build a candidate PMU set. Then we apply an unsupervised metric learning-based clustering method to this candidate set to label professional malicious users. To be specific, we apply the attention mechanism in metric learning to enhance the model. We conduct experiments on four real-world datasets: Amazon, Yelp, Taobao, and Jingdong. The results demonstrate that our proposed method can solve this unsupervised malicious user detection problem. Moreover, their performance of the state-of-the-art recommender models can be enhanced by taking MMD as a preprocessing stage.  
	
	We summarize the main contributions as follows.
	\begin{itemize}
		\item This is the first work focusing on solving the professional malicious user detection issue utilizing both users' ratings and reviews to enhance the state-of-the-art recommender systems.
		\item A novel multi-modal unsupervised method-MMD-is proposed to detect professional malicious users with the modified RNN and attention metric learning-based clustering.
		\item Extensive experiments are conducted on four real-world e-commerce datasets to verify our proposed method. Moreover, by filtering professional malicious users, some state-of-the-art models are enhanced. 
	\end{itemize}
	
	The remainder of the paper is organized as follows. We first provide some preliminaries in Section 2 and then elaborate on our proposed method in Section 3. We present and discuss experimental results in Section 4 and review related work in Section 5. Finally, we conclude this paper and discuss future directions in Section 6.
	
	\section{Preliminaries}
	This section provides motivations, basic definitions, and background to professional malicious user detection. 	
	\subsection{Motivations}
	The professional malicious user (PMU) is defined by two motivations: 1) the e-commerce websites judge the credit of retailers according to the good rating/review rate among all the ratings/reviews. So the professional malicious users utilize untruthful negative reviews and low ratings to threaten retailers for illegal profits. Meanwhile, if a user always gives a high proportion of negative reviews or low ratings groundlessly, the websites will treat the user as a malicious user with traditional outlier detection and punish him/her. However, PMU can use the masking strategy to control the proportion of negative reviews/low ratings and avoid the traditional detection of websites, which makes it a challenge to detect PMUs. 2) PMUs give fake ratings or reviews, which are difficult to distinguish from truthful ratings and reviews because PMUs utilize masking strategies to disguise themselves as normal users. This makes the existing recommendation models inaccurate and inefficient and leads to a bad recommendation. If we detect PMUs and filter the fake ratings and reviews, the performance of recommendation models should be improved.
	
	\subsection{Basic Definitions}
	In a recommender system, let $ U $ denote a set of $ m $ users $ U=\{u_1,u_2...u_m\} $, and $ I $ denote a set of $ n $ items $ I=\{i_1,i_2...i_n\} $. $ r_{ui} $ means the rating user $ u $ marked for item $ i $, as the entry of user-item matrix $ R_{m\times n} $. We build a review user-item matrix $ V_{n\times m} $ with $v_{ui}$ in the same way. For each user $u$ and item $i$, $p_u$ and $q_i$ denote their latent vector learned by embedding models.  
	
	In this paper, we focus on detecting professional malicious users who utilize masking strategies:\\ 
	\textbf{\textit{Definition 1-Professional Malicious Users (PMUs)}}: Malicious users who give fake ratings and negative reviews to make illegal profits and utilize masking strategies to avoid detections. 
	
	PMUs usually give fake ratings and negative reviews, alternatively, and keep them in a ``safe'' proportion to avoid being detected: 
	\begin{equation}
	\begin{aligned}
	\left| {R_u^\text{fa}} \right|/\left| {{R_u}} \right| \le {\theta ^\text{fa}};\left| {V^\text{ne}_{u}} \right|/\left| {{V_u}} \right| \le {\theta ^\text{ne}},u \in U
	\end{aligned}
	\label{eq1} 
	\end{equation}
	where $R_u^\text{fa}$ and $V_u^\text{ne}$ denote the fake ratings and negative reviews sets; $R_u$ and $V_u$ denote the whole ratings and reviews set of user $u$, and $\theta ^\text{ne}$ and $\theta^\text{fa} $ denote the thresholds for detection models. 
	
	In order to maximize their profits by avoiding detections, PMUs use masking strategies, which means they do not give $v^\text{ne}$ and $r^\text{fa}$ for an item $i$ at the same time. Instead, they usually give high ratings with negative reviews or fake ratings with positive reviews:
	\begin{equation}\
	\begin{aligned}
	\frac{{\left| {({r_{ui}} \in R_u^\text{fa},{v_{ui}} \notin V_u^\text{ne})||({r_{ui}} \notin R_u^\text{fa},{v_{ui}} \in V_u^\text{ne})} \right|}}{{\left| {{R_u} \cup {V_u}} \right|}} \ge {\theta ^\text{mu}},
	\end{aligned}
	\label{eq2} 
	\end{equation}
	where $\theta^\text{mu} $ is the threshold for PMUs. Finally, we formulate professional malicious user detection as follows:\\\textbf{\textit{Definition 2-Professional Malicious User Detection}}:
	Given user set $U$, item set $I$, rating set $R$, and review set $V$ as inputs, the object of professional malicious user detection is to filter the users with restrictions above, and output the PMU set $ U^{\text{mu}} $:
	\begin{equation}\
	\begin{aligned}
	{U^{\text{mu}}} = \text{Detect}(U,I,R,V); ~~\textbf{\text{s.t.}}~~ \text{Eq(1),~Eq(2)}.
	\end{aligned}
	\label{eq3} 
	\end{equation}
	\subsection{Hierarchical Dual-Attention RNN}
	To utilize the review for professional malicious user detections, we employ a state-of-the-art RNN model, Hierarchical Dual-Attention RNN (HDAN) \cite{D24xu2019slanderous,D26yang2016hierarchical}, to project the review into a sentiment score. The structure of HDAN is shown in Fig.\ref{MMD1}. HDAN calculates update gate $ug_t$, reset gate $re_t$ and temporary state $\widetilde h_{t-1}$ Eq.(\ref{eq4}), Eq.(\ref{eq5}), Eq.(\ref{eq6}): 
	\begin{figure}[tbp]
		\centering
		\includegraphics[width=0.9\columnwidth]{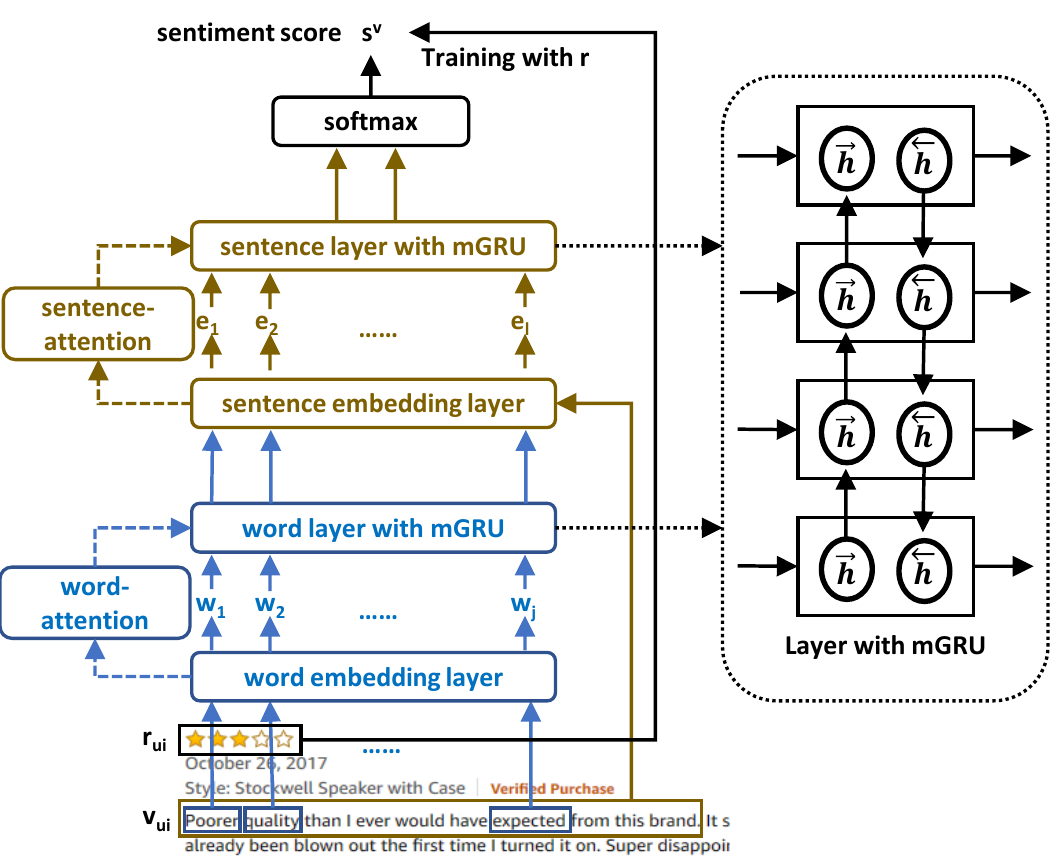}
		\caption{An example to project a review into a sentiment score. For each review $v_{ui}$, HDAN inputs word embedding $w$ and sentence embedding $e$ and ouputs a sentiment score $s^v$}
		\label{MMD1}
	\end{figure}
	
	\begin{equation}
	\begin{aligned}
	u{g_t} = \sigma ({W_{ug}}\widehat y + {U_{ug}}{h_{t - 1}} + {b_{ug}}),
	\end{aligned}
	\label{eq4} 
	\end{equation}
	\begin{equation}
	\begin{aligned}
	{\widetilde h_{t - 1}} = \tanh ({W_h}\widehat y + {re_t} \odot ({U_h}{h_{t - 1}}) + {b_h}),
	\end{aligned}
	\label{eq5} 
	\end{equation}
	\begin{equation}
	\begin{aligned}
	r{e_t} = \sigma ({W_{re}}\widehat y + {U_{re}}{h_{t - 1}} + {b_{re}}),
	\end{aligned}
	\label{eq6} 
	\end{equation}
	where $\widehat y$=$(y_{t-1}, y_t, y_{t+1})$ replaces $y_t$ in HAN for catching the sentiment in former $y_{t-1}$ and future state $y_{t+1}$. Finally, HDAN updates the information as follows:
	\begin{equation}
	\begin{aligned}
	{h_t} = (1 - u{g_t}) \odot {h_{t - 1}} + u{g_t} \odot {\widetilde h_{t - 1}},
	\end{aligned}
	\label{eq7} 
	\end{equation}
	
	Moreover, HDAN utilizes attention mechanisms to compute different weights for each word, as word-attention, and different weights for each sentence, as sentence-attention. In Fig.\ref{MMD1}, it is evident that the word ``\textit{Poorer}'' should take a more critical role than other words for this review. The details of attention computing are also discussed in \cite{D24xu2019slanderous}.
	
	In this paper, we take HDAN as a building block, which projects the review $v$ into a sentiment score $s^v$. To train HDAN, we use the ratings as the ground truth, and minimize the loss ${\text{L}_{v-r}}$ as follows:
	\begin{equation}
	\begin{aligned}
	{\text{L}_{v-r}} = \frac{1}{2}{\sum\limits_{U,I,V,R} {({r_{ui}} - s_{ui}^v)} ^2},
	\end{aligned}
	\label{eq8} 
	\end{equation}
	\begin{figure*}[tbp]
		\centering
		\includegraphics[width=1.7\columnwidth]{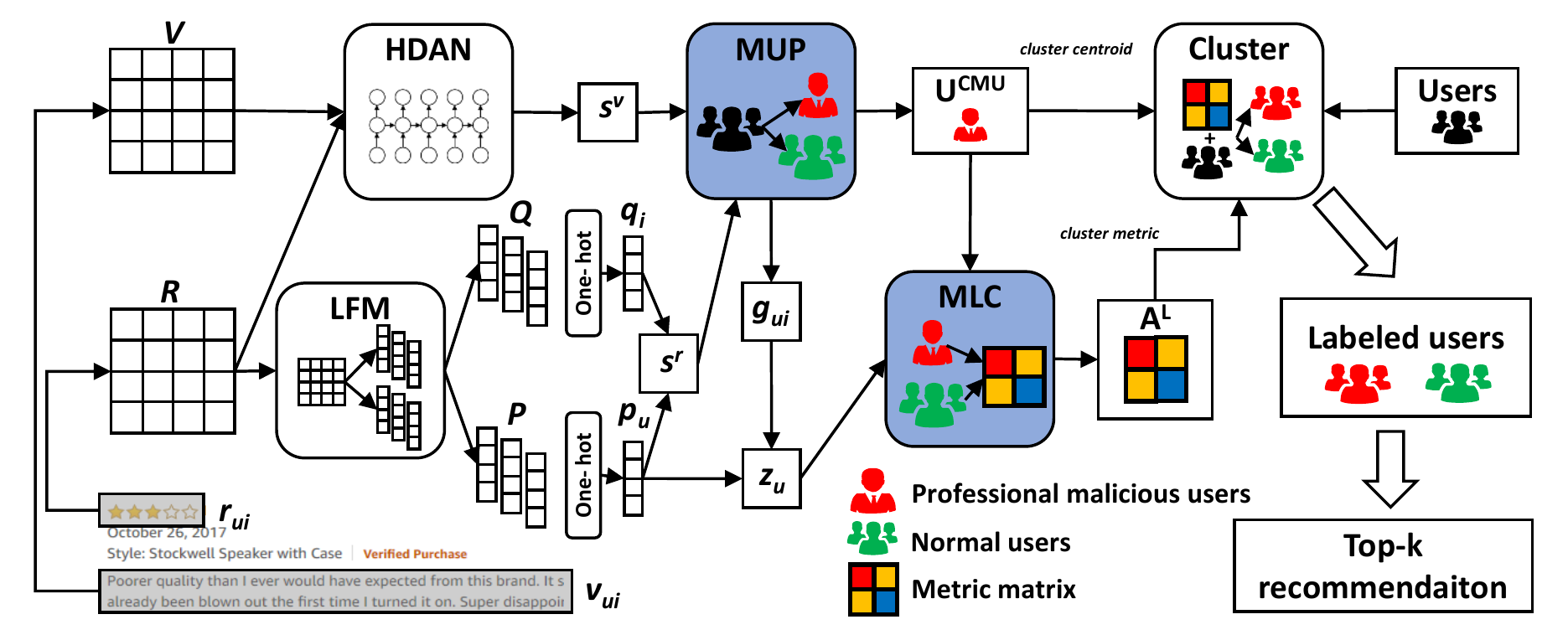}
		\caption{An illustration of the professional malicious user detection model (MMD). MMD comprises two important modules: professional malicious user profiling (MUP), attention metric learning (MLC)), which are encompassed by grey rectangles.}
		\label{MMD2}
	\end{figure*}
	\subsection{Metric Learning for clustering}
	The key to this detection is how to distinguish professional malicious users from normal users. The intuitive idea is to maximize the distance between malicious and normal users with clustering. In this paper, we utilize Metric Learning, a popular theory that is widely applied in clustering, embedding, and image recognition. The paradigm of metric learning is to estimate a proper ``distance'' between entities to measure the relationships \cite{D25xing2003distance}. Consider learning a distance metric matrix $A$ as follows: 
	\begin{equation}
	\begin{aligned}
	{d_\text{A}}({p_j},{p_k}) = {\left\| {{p_j} - {p_k}} \right\|_\text{A}} = \sqrt {{{({p_j} - {p_k})}^T}c^2A({p_j} - {p_k})},
	\end{aligned}
	\label{eq9} 
	\end{equation}
	where $j,k \in U$, $p_j$ and $p_k$ are latent vectors for $j,k$, and $c$ is a regular parameter. Note that matrix $A$ is a metric (it satisfies non-negativity and the triangle inequality in latent space) when $A \succeq 0$. To do clustering, metric learning attempts to learn a metric in which
	the different axes are given different ``weights''. 
	
	A simple idea of defining a criterion for a desired metric is to minimize the distance $d_\text{A}(j,k)$ if $j,k$ are in the same user subgroup $S$ (which means $j,k$ should be closer under the metric matrix $A$), and add some constraint to ensure $A$ does not force the user set into a point as a ``metric''. This gives an optimization problem:
	\begin{equation}
	\begin{aligned}
	\mathop {\min }\limits_A \sum\limits_{j,k \in S} {{d_\text{A}}({p_j},{p_k})}; 
	\end{aligned}
	\label{eq11} 
	\end{equation}
	\begin{equation}
	\begin{aligned}
	\textbf{s.t.}~~\sum\limits_{ j,k \in (U-S)} {{d_\text{A}}({p_j},{p_k})}  \ge c, A \succ 0. 
	\end{aligned}
	\label{eq12} 
	\end{equation}
	
	The method of solving this optimization is given in \cite{D25xing2003distance}, which is very clear to follow. However, for professional malicious user detection, there are two limitations for applying metric learning directly: 1) there is no dataset with PMU labels, which means that we cannot build the user subgroup $S$. And 2) $p_j$ and $p_k$ only contain the side information of user $j,k$, without considering the interactions (for example, masking strategies of PMUs) in recommender systems. To the best of our knowledge, MMD is the first model that combines HDAN and metric learning for PMU detection with reviews and ratings in recommender systems. Some important notations are shown in Table \ref{MDtable1}:
	\begin{table}[htbp]
		\centering
		\caption{Notation List}
		\begin{tabular}{c|l}
			\hline
			Notation & Description \\
			\hline
			$ U $ & user set \\
			$ I $ & item set \\
			$ R, V$ & rating/review set\\
			$ U^\text{mu}$ & PMU set\\
			$ U^\text{cmu}$ & candidate PMU set\\
			$ m, n$ & number of users/items\\
			$ r_{ui} $ & $ u $'s rating on item $ i $ \\
			$ v_{ui} $ & $ u $'s review on item $ i $ \\
			$ p_u, q_i $ & user $u$/item $i$'s latent vectors\\
			$ s^r $ & rating score ($p_u \bullet q_i$) \\
			$ s^v $ & review's sentiment score \\
			$ g_{ui} $ & sentiment gap between $s^r_{ui}$ and $s^v_{ui}$ \\
			$ k $ & the clustering number \\
			$ A_\text{L} $ & distance metric matrix \\
			$ \alpha^\text{g} $ & sentiment gap threshold  \\
			$ \theta^\text{mu} $ & detection threshold  \\
			\hline
		\end{tabular}%
		\label{MDtable1}%
	\end{table}%
	\vspace{-15pt}
	\section{Professional Malicious User Detection}
	In this section, we first present the professional malicious user profiling model (MUP), followed by the attention metric learning for clustering, MLC. Lastly, we analyze the time complexity of MMD.
	\subsection{Framework}
	To tackle the professional malicious user detection, we propose MMD, an unsupervised learning model, which applies metric learning and deep learning with both reviews and ratings. The framework of MMD is shown in Fig.\ref{MMD2}.
	
	At the beginning, MMD utilizes one-hot encoding to select user $u$ and item $i$, then projects them into $p_u$ and $q_i$. Latent factor model (LFM), which is the most widely used model in recommender area \cite{D27he2016fast}, is employed to get  $p_u$ and $q_i$, and calculates rating score $s^r$ with them. Meanwhile, we employ HDAN to project the review $v_{ui}$ to a sentiment score $s^v$. Note that we use ratings $r_{ui}$ as training ground truth for LFM and HDAN. Then we feed $s^v$ and $s^r$ into the professional malicious user profiling model (MUP). This model outputs the sentiment gap vector $g_{ui}$ and labels professional malicious users to build a candidate set $U^\text{cmu}$. We combine $g_{ui}$ and $p_u$ to build a profile vector $z_u$ and utilize $U^\text{cmu}$ as ground truth to learn a proper metric matrix $A_\text{L}$ for professional malicious user detection. Especially, we apply attention to metric learning to enhance the model. Finally, we choose $U^\text{cmu}$ as cluster centroids, $A_\text{L}$ as the clustering metric to cluster the users, which achieves the professional malicious users $U^\text{mu}$. The details of MMD are introduced in the following subsections.
	\subsection{Professional Malicious User Profiling (MUP) model}
	
	To profile professional malicious users, we need to analyze their masking strategies. Unlike normal users, professional malicious users always use the following two interactions as masking strategies: 1) giving a high rating $r_{ui}$ with a negative review $v^\text{ne}_{ui}$; 2) giving a positive review $v_{ui}$ with a fake rating $r^\text{fa}_{ui}$. By utilizing the masking strategies, professional malicious users can avoid statistic outlier detection with thresholds $\theta^r, \theta^v$, and confuse the recommender system.
	
	Taking a deep insight, we notice that there always exist sentiment gaps between each PMU's ratings and reviews, which are the most remarkable differences from a normal user's actions. So we first utilize HDAN to project review $v$ onto a sentiment score $s^{v}_{ui}$ (Eq.(\ref{eq13})). Meanwhile we embed users and items onto latent space $P,Q$ with basic LFM (Eq.(\ref{eq29})), and calculate a rating score $s^{r}_{ui}$ with $p_u,q_i$ (Eq.(\ref{eq14})).
	
	Specifically, we minimize $\text{L}_\text{LFM}$ to achieve latent vetor $p_u$ and $q_i$ with rating $r_{ui}$ as input:
	\begin{equation}
	\begin{aligned}
	{\text{L}_\text{LFM}} = \sum\limits_{u \in U,i \in I} {{{({p_u}^\text{T}{q_i} - {r_{ui}})}^2}}.
	\end{aligned}
	\label{eq29} 
	\end{equation}
	\begin{equation}
	\begin{aligned}
	s_{ui}^v = \text{HDAN}({v_{ui}}).
	\end{aligned}
	\label{eq13} 
	\end{equation}
	\begin{equation}
	\begin{aligned}
	s_{ui}^r = {p_u} \odot {q_i}.
	\end{aligned}
	\label{eq14} 
	\end{equation}
	
	Note that $s^{v}_{ui}$ is in the same range of $s^{r}_{ui}$, which is 1 to 5. Hence, we utilize the sentiment gap $g_{u}$ to profile the users. Specifically, for each user-item pair ($u,i$), we calculate sentiment gap $g_{ui}$. 
	\begin{equation}
	\begin{aligned}
	g_{ui}=|s^{r}_{ui}-s^{v}_{ui}|_\text{abs}
	\end{aligned}
	\label{eq15} 
	\end{equation}
	where $|a|_\text{abs}$ means the absolute value of $a$. For $u$ as a normal user, the gap $g_{ui}$ should be small (threshold $\alpha^\text{g}$) because no matter how reviews and ratings are given, they all contain users' real sentiment for items. In comparison, for $u$ as a professional malicious user, the gap should be huge for most items (threshold $\theta^\text{mu}$) because of their masking strategies. We build gap vector $\hat{g_u}$ for each user $u$ who has $k$ items with feedbacks, which entries are $g_{uk}$:
	\begin{equation}
	\begin{aligned}
	{\hat{g_u}} = \{ {g_{u1}},{g_{u2}}......,{g_{u(k - 1)}},{g_{uk}}\} ,k \in I.\
	\end{aligned}
	\label{eq16} 
	\end{equation}
	
	To this end, we profile professional malicious user with the following practicable rule, label them and build a candidate professional malicious user set $U^\text{cmu}$:
	\begin{equation}
	\begin{aligned}
	u \in {U^\text{cmu}};
	\end{aligned}
	\label{eq17} 
	\end{equation}
	\begin{equation}
	\begin{aligned}
	\textbf{s.t.}~~\frac{{\left| {\left\{ {{g_{ui}}\left| {{g_{ui}} \ge {\alpha ^\text{g}}} \right.} \right\}} \right|}}{{\left| {\hat{g_u}} \right|}} \ge {\theta ^\text{mu}};{g_{ui}} \in {\hat{g_u}},u \in U,i \in I
	\end{aligned}
	\label{eq18} 
	\end{equation}
	
	With MUP, we can label professional malicious users. However, because the amount of professional malicious users only takes small portions of users, it is still challenging to use these labeled professional malicious users directly to learn more information, discover the insight, and improve the recommendation. To tackle this issue, we treat all these users in $U^\text{cmu}$ as similarities, which means they can be categorized into the same cluster in a specific latent metric space. In this latent metric space, we can learn more about professional malicious users and normal users. In the next section, we will introduce how our method, attention metric learning (MLC), learns this specific latent metric space. 
	\subsection{Attention Metric Learning for Clustering (MLC)} 
	\subsubsection{Model Construction}
	To learn a specific metric space, we do clustering to find the inner connections between professional malicious users. Towards a comprehensive understanding, we consider that professional malicious users are different from normal users in two perspectives: first, they have different attributes. Professional malicious users are signed up for their particular purpose, which is different from normal users. Second, there is a noticeable sentiment gap between ratings and reviews for professional malicious users, while for normal users, the sentiment gap is small.  Without loss of generality, we combine users' p-dimension latent vector $p_u$ and k-dimension gap vector $\hat{g_u}$ to construct the (p+k)-dimension profile vector $z_u$, which contains the information of users' attributes and sentiment gaps. Then we apply metric learning to learn the proper latent metric with $U^\text{cmu}$.
	\begin{equation}
	\begin{aligned}
	{z_u} = {p_u} \oplus \hat{g_u},
	\end{aligned}
	\label{eq19} 
	\end{equation}
	where $\oplus$ means a direct combination. With this representation vector, we first rewrite Eq.(\ref{eq9}) as follows: 
	\begin{equation}
	\begin{aligned}
	{d_\text{A}}({z_j},{z_k}) = {\left\| {{z_j} - {z_k}} \right\|_\text{A}} = \sqrt {{{({z_j} - {z_k})}^T}c^2A({z_j} - {z_k})},
	\end{aligned}
	\label{eq20} 
	\end{equation}
	where $j,k \in U$. If we utilize Eq.(\ref{eq20}) directly, some critical information may be ignored, which leads to learning inaccurate metric. In real-world recommender systems, to cluster different users, the different attribute should take different importance and arrange different weights, which is the theory of attention mechanism \cite{D28vaswani2017attention}. The intuitive idea is that the attributes and sentiment gaps should make different contributions to metric learning. With this restriction, we can add the attention vector $t$ into Eq.(\ref{eq20}), using $t \otimes  z$ to replace $z$. Note that $\otimes$ means element-wise product. 
	
	There are various ways to define attention vectors. Specifically, in our situation, it is a local optimization issue, where the general attention vector can achieve a proper performance without huge additional computing cost \cite{D29luong2015effective}. Hence, we utilize the general attention style to compute the attention vector $t$ as follows:
	\begin{equation}
	\begin{aligned}
	\text{f}({z_u},A) = z_u^T{W_t}A,
	\end{aligned}
	\label{eq21} 
	\end{equation}
	\begin{equation}
	\begin{aligned}
	{t_u} = \text{align}({z_u},A) = \frac{{\exp (f({z_u},A))}}{{\sum\nolimits_\text{A} {\exp (f({z_u},A))} }},
	\end{aligned}
	\label{eq22} 
	\end{equation}
	where $W_t$ is the general weights for attention $t$. Note that we utilize the attention vector to build a bridge between profile vector $z_u$ and metric $A$. Moreover, we take a deep insight into the form of $A$. $A$ is a (p+k) square metric matrix, where each entry of $A$ stands for a weight of the metric in this dimension. In our proposed model, we restrict the metric for attributes $p_u$ to be Euclidean distance, which means the entries of A should be initialized as follows:
	\begin{equation}
	\begin{aligned}
	{a_{i,j}} = \left\{ {\begin{array}{*{20}{c}}
		{0,i \ne j;}\\
		{1,i = j.}
		\end{array}} \right.i,j \in \left[ {1,p} \right].
	\end{aligned}
	\label{eq23} 
	\end{equation}
	
	While for the sentiment gap $\hat g_u$, we initialize the metric weight entries in $A$ with standard normal distribution $N(0,1)$. Note that our original metric matrix $A$ is partly diagonal, it can be learned quickly in the first p dimension, which is similar to \cite{D30he2019fast}. Moreover, to measure the relationship between users' attributes and sentiment gap, we also learn two p$\times$k matrices, which locate at the up-right and bottom-left of metric $A$, respectively. 
	\begin{figure}[tbp]
		\centering
		\includegraphics[width=0.9\columnwidth]{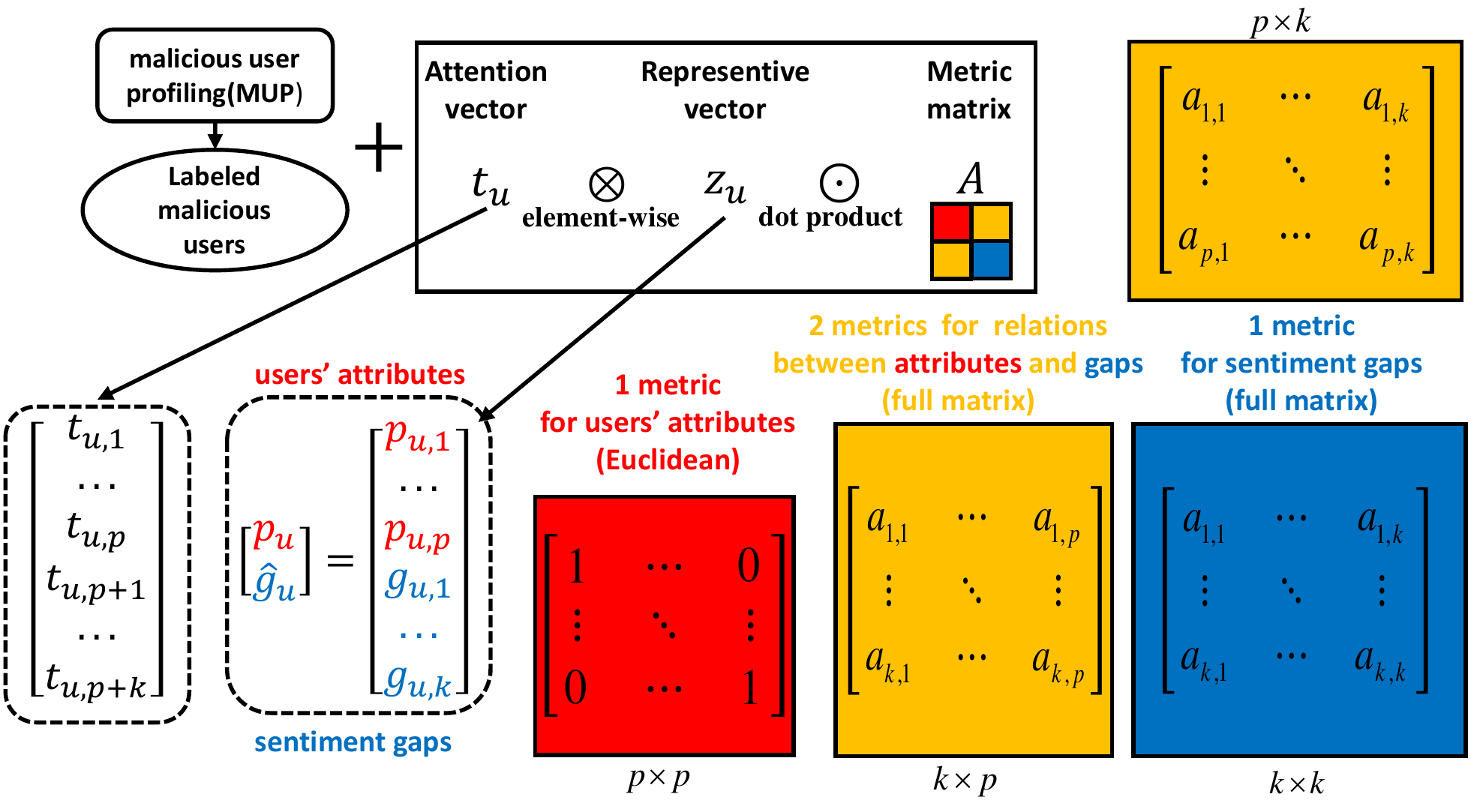}
		\caption{An illustration of attention metric learning for clustering.}
		\label{MMD3}
	\end{figure}
	
	To learn this metric matrix $A$$_\text{L}$, we need to build an objective function as Eq.(\ref{eq11}),(\ref{eq12}). To simplify the objective function, we jointly learn the metric $A$$_\text{L}$ and attention vector $t$ at the same time, with the following loss function:
	\begin{equation}
	\begin{aligned}
	{\text{L}_\text{MLC}} = \sum\limits_{j,k \in {U^\text{cmu}}\hfill\atop
		j,k' \notin {U^\text{cmu}}\hfill}^{U,I,R,V} {\left( {\lambda {d_\text{A}}({z_j},{z_k}) - (1 - \lambda ){d_\text{A}}({z_j},{z_{k'}}) + c} \right)} ,
	\end{aligned}
	\label{eq24} 
	\end{equation}
	where the $c$ is the parameter for normalization, which is the same as Eq.(\ref{eq11}). This function borrows the idea of BPR \cite{D31rendle2009bpr}, which maximizes the distance between different clusters (${ - (1 - \lambda ){d_\text{A}}({z_j},{z_{k'}})}, j,k' \notin U^\text{cmu}$) and minimizes the same cluster pairs (${\lambda{d_\text{A}}({z_j},{z_{k}})}, j,k \in U^\text{cmu}$). $\lambda$ is a parameter to tune the importance of different samples. 
	
	Without loss of generality, this loss function can be applied to learn different metrics. If we set $A$ to be a diagonal matrix, $a_{j,k}$=1, if $j=k$ while $a_{j,k}$=0 if not. This loss function fades to a Euclidean distance learning without attention vector. By arranging different forms of metric $A$, we can get the inner sight of the differences between professional malicious users and normal users.
	\subsubsection{Model Optimization}
	To make MLC less sensitive to the negative sampling, we also consider the restriction to the parameters in addition to minimizing the objective loss function $\text{L}_\text{MLC}$. Let $\Theta $ be the model parameters, which includes metric matrix $A$ and attention vector set $t$. Hence, we define the optimization objective for MLC as:
	\begin{equation}
	\begin{aligned}
	{\Theta ^*} = \arg \mathop {\min }\limits_\Theta  {\text{L}_\text{MLC}} + {\left\| \Theta  \right\|_2},
	\end{aligned}
	\label{eq25} 
	\end{equation}
	where ${\left\| \Theta  \right\|_2}$ is the regularization with F2-norm. In this formulation, we jointly learn metric matrix $A$ and attention vector set $t$ at the same time. Note that in real-world, PMU takes only small portion of users (nearly 10\%). We need to balance the weights of labeled (PMUs) and unlabeled users (normal users), which means that $\lambda$ should be larger than 0.5.
	
	\subsubsection{Learning Algorithm}
	This step updates model parameters by minimizing Eq.(\ref{eq24}). By utilizing this objective function, metric A and attention weight $W_t$ (Eq.(\ref{eq21})) are learned at the same time. It is a typical conventional minimization problem and can be approached with gradient descent. Specifically, we perform a gradient step for each involved parameter:
	\begin{equation}
	\begin{aligned}
	\Theta  = \Theta  - \eta \frac{{\partial {\text{L}_\text{MLC}}}}{{\partial \Theta }},
	\end{aligned}
	\label{eq26} 
	\end{equation}
	where $\Theta$=$\{$$A$$,W_t\}$. $\eta$ denotes the learning rate, which is parameter-dependent if some auto-adaptive SGD models are applied. In our proposed model, we set Adagrad \cite{D32duchi2011adaptive} as our SGD method. We can 1) sample the labeled professional malicious users repeatedly to build more samples (for each labeled professional malicious user, select 5 to 10 times unlabeled normal users to learn the metric); 2) enhance the importance of labeled users (a large $\lambda$). To validate the performance, we can monitor the return on a holdout validation dataset (which is the original data in Fig. \ref{MMD2}). In this way, we can achieve a learned metric matrix $A_\text{L}$.
	\subsubsection{Detect professional malicious users with metric A$_L$}
	After we learn a metric $A_\text{L}$, we do simple k-means clustering (actually, it is a 2-means clustering, which puts users into normal or professional malicious user set) to detect professional malicious users in original data. Specifically, we take all the professional malicious users in $U^\text{cmu}$ into the original data and do clustering in $A_\text{L}$ latent data space, and label all the users in $U$. This k-means model can achieve convergence rapidly because we give some labeled users as heuristic information. With this step, we can get the professional malicious user set $U^\text{mu}$, which is the cluster with more labeled professional malicious users. Lastly, we conclude the processing of MMD in Algorithm 1.
	\begin{algorithm}
		\caption{Attention Metric learning for professional Malicious user Detection(MMD)}
		\LinesNumbered 
		\KwIn{Users $U$, items $I$, ratings $R$, reviews $V$, sentiment gap $\alpha^\text{g}$, detection threshold $\theta^\text{mu}$, learning rate $\eta$, hyperparameter $O$ for HDAN, MUP and MLC.}
		\KwOut{Professional malicious users $U^\text{mu}$, distance metric $A_\text{L}$.}
		Initialize $O$, distance metric $A$ with Eq.(\ref{eq23})\;
		\textbf{Professional Malicious User Profiling:(line 2-6)}\\
		~~~~~Calculate $s^v$ with HDAN (Eq.(\ref{eq13}))\;
		~~~~~Calculate $s^r$ with LFM (Eq.(\ref{eq14}), Eq.(\ref{eq29}))\;		
		~~~~~Calculate $\hat g_u$\;
		~~~~~Label the candidate professional malicious user $U^\text{cmu}$ with $\alpha^\text{g}, \theta^\text{mu}$\;
		\textbf{MLC:(line 7-17)}\\
		~~~~~Build $z^u$ with Eq.(\ref{eq19})\;
		\While{not converge}{
			Randomly draw an example ($j,k,k'$) from $ U $\;
			Calculate attention vector $t_u$ with Eq.(\ref{eq21}),(\ref{eq22})\;
			Calculate $\text{L}_\text{MLC}$ with Eq.(\ref{eq24})\;
			Update $A$, $W_t$:\\
			~~$A  \leftarrow  A - \eta \frac{{\partial {\text{L}_\text{MLC}}}}{{\partial A}}$\;
			~~$W_t \leftarrow W_t - \eta \frac{{\partial {\text{L}_\text{MLC}}}}{{\partial W_t}}$\;
		}
		~~~~~\textbf{Return} $A_\text{L}$\;
		K-means Clustering with $U^\text{cmu}$, $A_\text{L}$\;
		Label users in $U$ as $U^\text{mu}$\;
		\textbf{Return} $U^\text{mu}$.
	\end{algorithm}
	\subsection{Time Complex Analysis}
	
	In MMD, there are three sub-modules: MUP, MLC and K-means. Note that these models are employed sequentially in MMD, so the time complexity of MMD should be: O$_{\text{MMD}}$=O$_{\text{Kmeans}}$+O$_{\text{MUP}}$+O$_{\text{MLC}}$. For k-means, the time complexity is O$(n \times k \times It)\approx $O$(n\text{log}n)$, where $n$ is the data scale, $k$ is 2 in MMD for k-means, and $It$ is the iteration times. For MUP, which consists of HDAN and LFM, the time complexity is: $O_{\text{MUP}}$=$O_{\text{HDAN}}$+$O_{\text{LFM}}$. However, the output of LFM is fixed in our model, which means it could be pretrained as preprocessing. So O$_{\text{MUP}}\approx  $O$_{\text{HDAN}}=$O$(nd^2)$, where $d$ is the dimensions of input vectors. For MLC, let O$_\text{MLC}$ denote the time of learning A. Note that with different form settings of $A$, the time complexity is different. If we set $A$ as a diagonal matrix, O$_\text{MLC}$ should be O$ ((k+p)log(k+p)) $, where A is a $k+p$ dimension matrix. If we set $A$ as a full matrix, it adds to O$((k+p)^2)$. Since we define the form of $A$ as Fig.\ref{MMD3}, O$_\text{MLC}$ is O$(p\log p + {k^2})$.
	
	Because of the sequential process of MMD, the whole time complexity should be: 
	\begin{equation}
		\begin{aligned}
		\text{O}_{\text{MMD}}=\text{O}(n\text{log}n)+\text{O}(nd^2)+\text{O}(p\log p + {k^2}).
		\end{aligned}
		\label{eq30} 
	\end{equation}
	Utilizing parallel processing or other computing frameworks may accelerate our model, where we leave as an important future work.

	\section{Experiments}
	In this section, we conduct extensive experiments to answer the following issues:
	
	\textbf{RQ1:} Can our proposed method MMD outperform the state-of-the-art malicious user detection models?
	
	\textbf{RQ2:} How is the effect of metric learning when it is applied in the detection, and can it help to relieve the lacking of labeled professional malicious users?
	
	\textbf{RQ3:} How do the key parameters, such as $\alpha$, $\theta$, affect the detection performance?
	
	\textbf{RQ4:} How does MMD benefit the recommender system with the detected professional malicious users?
	\subsection{Experimental Settings}
	\subsubsection{Data Descriptions}
	We conduct abundant experiments on Amazon.com dataset\footnote[1]{https://jmcauley.ucsd.edu/data/amazon} and Yelp for RecSys\footnote[2]{https://www.kaggle.com/c/yelp-recsys-2013}. Amazon and Yelp datasets are two public datasets with abundant textual reviews and ratings. Moreover, we also collect two real-world datasets from Taobao\footnote[3]{https://www.taobao.com} and Jindong\footnote[4]{https://www.jd.com} to validate MMD. These datasets all contain ratings in the range of 1 to 5. The details of datasets are shown in Table \ref{MDtable2} (\textit{/s, /r, /u} mean per sentence/review/user). 
	
	Since the original data of Amazon and Yelp are vast and sparse, we sample a small subset of the data to validate our method. Note that in the real world, the PMU ratio is about 10\%. Specifically, we randomly select 450 users with more than five feedbacks (reviews and ratings). Note that Amazon and Yelp are standard datasets without professional malicious users. To validate our MMD, we add 50 artificial professional malicious users with negative feedbacks on random items (note that these artificial malicious users employ the masking strategies). For Taobao and Jindong, we select 450 normal users and 50 true professional malicious users, which are listed on a website (www.taocece.com, where the electronic retailers upload the professional malicious users' IDs to this website).
	\begin{table}[h]
		\centering
		\caption{Datasets' Characteristics }
		\label{MDtable2}
		\resizebox{0.49\textwidth}{!}{
			\begin{tabular}{lllll}
				\hline\hline
				Dataset     & Amazon  & Yelp & Taobao & Jingdong
				\\ \hline
				\hline
				\#user    & 30,759  & 45,980 & 10,121 & 8,031
				\\ \hline
				\#item    & 16,515  & 11,537  & 9,892 & 3,025
				\\ \hline
				\#review    & 285,644  & 229,900 & 10,791 & 8,310
				\\ \hline
				\#rating    & 285,644  & 229,900 & 49,053 & 25,152
				\\ \hline
				Sparsity    & 0.051\%  & 0.043\% & 0.049\% & 0.12\%
				\\ \hline
				PMU ratio    & 0\%  & 0\% & 9.31\% & 10.71\%
				\\ \hline
				PMU fake ratings/ratings& 0\%  & 0\% & 45.5\% & 56.7\%
				\\ \hline
				PMU fake reviews/reviews& 0\%  & 0\% & 66.6\% & 54.5\%
				\\ \hline
				Avg words \textit{/s} &10.1 &9.9 & 12.7 & 13.2
				\\ \hline
				Avg words \textit{/r} &104 &130 & 65 & 70
				\\ \hline
				Avg sentences \textit{/r} &9.7 &11.9 & 4.9 & 5.1
				\\ \hline
				Avg reviews \textit{/u} &9.29 &5.00 & 1.06 & 1.03
				\\ \hline\hline
		\end{tabular}}
	\end{table}
	
	\subsubsection{Performance Evaluation}
	Following the prominent work in malicious user detection \cite{D24xu2019slanderous} and shilling attack detection \cite{D33wu2012hysad}, we evaluate our proposed detection model with objective and subjective evaluations.
	
	\textbf{a) Objective evaluation}: we employ \textit{specificity} and \textit{sensitivity} as the objective metrics \cite{D34cao2013shilling}:
	\begin{equation}
	\begin{aligned}
	SEN = \frac{{\# \text{true~positive}}}{{\# \text{true~positive} + \# \text{false~negative}}},
	\end{aligned}
	\label{eq27} 
	\end{equation}
	\begin{equation}
	\begin{aligned}
	SPE = \frac{{\# \text{true~negative}}}{{\# \text{true~negative} + \# \text{false~positive}}}.
	\end{aligned}
	\label{eq28} 
	\end{equation}
	To explicitly introduce, we give definitions in Table \ref{MDtable3}, where the \textit{specificity} ($ SPE $) measures the proportion of correctly normal users,
	and the \textit{sensitivity} ($ SEN $) measures the proportion of correctly detected labeled professional malicious users. We also utilize $F-score$= $\frac{{2 \times SEN \times SPE}}{{SEN + SPE}}$ to balance $SEN$ and $SPE$.  
	\begin{table}[h]
		\centering
		\caption{Definitions for \textit{specificity} and \textit{sensitivity} }
		\label{MDtable3}
		\resizebox{0.45\textwidth}{!}{
			\begin{tabular}{|l|l|l|}
				\hline
				\diagbox{Actual}{Detected} & Malicious Users  & Normal Users
				\\ \hline
				Malicious Users    & true~positive  & false~negative
				\\ \hline
				Normal Users    & false~positive  & true~negative
				\\ \hline
		\end{tabular}}
	\end{table}
	
	Moreover, to evaluate the enhancement of recommender systems, we measure the quality of recommendation with \textit{Hit Ratio} (HR) and \textit{Normalized Discounted Cumulative Gain}. Specifically, HR@N is a metric to measure whether the testing item exists in the Top-N recommendation list, where 1 for yes and 0 for no; NDCG@N measures the position of the testing item in the top-N list, the higher, the better. The default setting of N is 5 without special mention. Both metrics are commonly applied for evaluating the recommender systems.
	
	\textbf{b) Subjective evaluation}: we employ 20 students to distinguish the detection results. After we achieve the professional malicious user set $U^\text{mu}$, we ask these students to label these users (1 for professional malicious users, 0 for normal users). We treat the students' results as the ground-truth and evaluate the detection model with the comparisons, which is also a supplement for subjective evaluation.
	\subsubsection{Baseline Methods}
	For professional malicious user detection, an unsupervised detection problem, there are few types of research available in this area. So we compare MMD with the following methods. Specifically, there are two unsupervised methods and two supervised methods. 
	
	\textbf{K-means++ Clustering} is a basic unsupervised method that clusters the data into k clusters. In this paper, we treat users as only normal or malicious ones, which means it is a 2-clustering issue, and we label the cluster with more labeled users as malicious users. \textbf{Statistic Outlier detection (SOD)} is a basic statistic method, which counts the negative feedbacks of each user, and labels the users with a high percentage of negative feedbacks as malicious users. This method is sensitive to the negative feedback threshold $\theta$. \textbf{Hy-sad} \cite{D33wu2012hysad} is a supervised hybrid shilling attack detection method, which introduces MC-Relief to select useful detection metrics. We find that using the user's embeddings $p_u$ learned by LFM leads to better performance, so we report this specific setting. \textbf{Semi-sad} \cite{D34cao2013shilling} is a semi-supervised learning based shilling attack detection algorithm, which tackles the labeled and unlabeled data at the same time. \textbf{CNN-sad} \cite{46tong2018shilling} is a novel convolutional neural network-based method, which applies a transformed network structure to exploit deep-level features from users rating profiles. CNN-SAD can detect shilling attacks more efficiently, which can be adapted for PMU detection. \textbf{SDRS} \cite{D24xu2019slanderous} utilizes a dual-attention RNN and a modified GRU to compute an opinion level for reviews, then a joint filtering method is proposed to detect malicious users. 
	
	Note that these baselines, especially two state-of-the-art methods (Hy-sad and Semi-sad), are validated only in the standard datasets with only ratings, which can not utilize the abundant information hidden in reviews to detect professional malicious users. In comparison, our proposed MMD can tackle ratings and reviews at the same time, which is an improvement in this area. Moreover, the practical application in real-world scenarios should also be explored, and we will do this valuable work in the following subsections.
	
	\subsubsection{Parameter Settings}
	To explore the hyper-parameter space for all methods, we randomly holdout a training interaction for each user as the validation set. First, for all the baselines, we report the best results to make a fair comparison. Specifically, for K-means++, we set \textit{k}=2, and initialize original cluster centroids from labeled professional malicious users and normal users with clustering user's attributes $p_u$. For statistic outlier detection, we set negative feedback threshold $\theta$=0.8. For Hy-sad and Semi-sad, we utilize labeled PMUs (50 PMUs in 500 users) to train the models because these models are supervised. Note that the labels of professional malicious users are difficult to obtain. We tune the size of labeled professional malicious user sets with an upper bound 10\% of the dataset. Moreover, in Taobao and Jingdong datasets, we do not inject artificial professional malicious users to simulate the application scenario in the real-world.
	
	For MMD, we initialize LFM and HDAN by \cite{D24xu2019slanderous}. Specifically, we fix the embedding dimension as 32 for users and items, then tune other parameters as follows. In MLC, we set $\lambda$=0.6. In professional malicious user profiling, we set the sentiment gap threshold $\alpha^\text{g}$=3.5, and detection threshold $\theta^\text{mu}$=0.7. We use these default settings if there are no additional instructions.
	
	\subsection{Performance Comparison (RQ1)}
	Here we compare the performance of MMD with baselines. We explore the detection results with different datasets. The results are listed in Table \ref{MDtable4}, Table \ref{MDtable5}. Inspecting the results from top to bottom, we have the following observations.
	\begin{table}[h]
		\centering
		\caption{PMU detection with labeled artificial professional malicious users in Amazon and Yelp. \textbf{bold} stands for MMD and $^*$ marks the best performance.}
		\label{MDtable4}
		\resizebox{0.49\textwidth}{!}{
			\begin{tabular}{llllllll}
				\hline\hline
				&    \multicolumn{3}{c}{Amazon} & \multicolumn{3}{c}{Yelp}
				\\ \hline
				& \textit{SEN}  & \textit{SPE}  &F-score& \textit{SEN} & \textit{SPE}&F-score
				\\ \hline
				K-means++    & 0.381  & 0.706&0.494& 0.201 & 0.773&0.318
				\\ \hline
				SOD    & 0.062  & 0.984 &0.113& 0.041 & 0.962&0.076
				\\ \hline
				Hy-sad    & 0.371  & 0.853&0.516 & 0.542 & 0.889&0.671
				\\ \hline
				Semi-sad  &0.442 &0.784 &0.563& 0.661 & 0.933&0.773
				\\ \hline
				CNN-sad  &0.552 &0.791 &0.650& 0.663 & 0.922&0.771
				\\ \hline
				SDRS  &0.651 &0.874 &0.745& 0.764 & 0.913&0.831
				\\ \hline
				MLC &0.861 &0.967 &0.910& 0.821 & 0.938&0.874
				\\ \hline
				MUP &0.662&1\textbf{*} &0.795& 0.742 & 1\textbf{*}&0.850
				\\ \hline
				MMD\\(MUP+MLC) &\textbf{0.921*} &\textbf{0.970} &\textbf{0.944*}& \textbf{0.98*} & \textbf{0.996}&\textbf{0.987*}
				\\ \hline\hline
		\end{tabular}}
	\end{table}
	\begin{table}[h]
		\centering
		\caption{PMU detection without labeled artificial professional malicious users in Taobao and Jindong. \textbf{bold} stands for MMD and $^*$ marks the best performance.}
		\label{MDtable5}
		\resizebox{0.49\textwidth}{!}{
			\begin{tabular}{lllllll}
				\hline\hline
				&    \multicolumn{3}{c}{Taobao} & \multicolumn{3}{c}{Jingdong}
				\\ \hline
				& \textit{SEN}  & \textit{SPE}  &F-score& \textit{SEN} & \textit{SPE}&F-score
				\\ \hline
				K-means++    & 0.141  & 0.728&0.234& 0.44 & 0.667&0.530
				\\ \hline
				SOD    & 0.022  & 0.978 &0.039& 0.14 & 0.896&0.242
				\\ \hline
				Hy-sad    & -  & -&- & - & -&-
				\\ \hline
				Semi-sad  &- &- &-& - & -&-
				\\ \hline
				CNN-sad  &- &- &-& - & -&-
				\\ \hline
				SDRS  &0.451 &0.884 &0.597& 0.732 & 0.911&0.811
				\\ \hline
				MLC &- &- &-& - & -&-
				\\ \hline
				MUP &0.641&0.996\textbf{*} &0.779& 0.62 & 0.998\textbf{*}&0.764
				\\ \hline
				MMD\\(MUP+MLC)  &\textbf{0.941*} &\textbf{0.987} &\textbf{0.963*}& \textbf{0.96} & \textbf{0.989}&\textbf{0.974*}
				\\ \hline\hline
		\end{tabular}}
	\end{table}
	
	First, on Amazon and Yelp with artificial labeled professional malicious users, supervised baseline models (Hy-sad, Semi-sad, CNN-sad, and SDRS) largely outperform the unsupervised baselines (K-means++ and SOD) on F-score (Table \ref{MDtable4}). While on Taobao and Jingdong, where the datasets are without labeled malicious, supervised models, such as Hy-sad, Semi-sad, and MLC, do not work at all. The result demonstrates the positive effects of labeled data for detections, also the negative effects of narrow applications for supervised models.
	
	Second, among all the supervised models, MLC consistently outperforms the other models. The enhancement of MLC demonstrates that the performance of a supervised model can be significantly improved by cooperating with metric learning. Because MLC utilizes metric learning and attention at the same time, we speculate that there exist complex relationships that can not be measured by simple metrics. Our method can catch this kind of relationship. Among all the unsupervised models, MUP consistently outperforms the other models, K-means++ and SOD. The enhancement of MUP demonstrates that MUP can obtain the characteristics of professional malicious users and achieves a better performance.
	
	Third, on all four datasets, our proposed model MMD (MLC+MUP) outperforms all the baselines in terms of F-score. Note that MMD is a combination of MLC and MUP, which ensures its superiority over supervised and unsupervised models in general. On different application scenarios (with or without labeled professional malicious users), MMD can achieve a satisfying result, where the improvement over Semi-sad in Amazon/Yelp is 67.6\%/34.6\%, over K-means++ in Taobao/Yelp is 311\%/302\%. The result justifies the positive effect of our MMD on learning better metric representations for professional malicious user detection, thus leads to better detection performance.
	
	Finally, taking a deep insight into the results, we notice some interesting phenomena. Among all the datasets on all metrics, SOD, a widely-applied simple outlier detection model, achieves the worst performance, which gives evidence that the effect of masking strategies can avoid traditional detections. Moreover, note that the improvement of MMD over MUP, MMD over MLC, is smaller than those over other baselines. Then the reason why we combine MLC and MMD to build MMD is that: first, MLC can not be applied to unlabeled datasets because it is a supervised model; second, MUP does not perform well in $ SEN $ among all baselines, which means that MUD may treat some professional malicious users as normal users, which leads to a high $ SPE $ and low $ SEN $. With the combination, MMD can achieve superior performance on different application scenarios and avoid wrong detections.
	
	Without the loss of generality, we randomly select 50 users in Taobao (25 professional malicious users reported by the website, 25 normal users), and employ 20 students to label the dataset (These students do not know the distribution of this dataset.). The results are reported in Table \ref{MDtable6}. Note that the students label 24 true positive and 24 true negative users on average. In comparison, MMD labels 23 true positive and 24 true negative users, which achieves the same level with students and surpasses baselines.
	\begin{table}[h]
		\centering
		\caption{Comparison with students' subjective detections. \textbf{bold} stands for MMD and $^*$ marks the best performance.}
		\label{MDtable6}
		\begin{tabular}{llll}
			\hline\hline
			&    \multicolumn{3}{c}{Taobao} 
			\\ \hline
			& \textit{SEN}  & \textit{SPE}  &F-score
			\\ \hline
			K-means++    & 0.52  & 0.6&0.557
			\\ \hline
			Semi-sad  &0.68 &0.76 &0.718
			\\ \hline
			MLC &0.92 &0.88 &0.899
			\\ \hline
			MUP &0.76&1\textbf{*} &0.867
			\\ \hline
			MMD\\(MUP+MLC)  &\textbf{0.92} &\textbf{0.96} &\textbf{0.94}
			\\ \hline
			Students &0.96\textbf{*} &0.96 &0.96\textbf{*}
			\\ \hline\hline
		\end{tabular}
	\end{table}
	
	To be specific, we give some validations here, to reveal a real example in Taobao. We list some PMUs detected by MMD in Fig.\ref{MMD4-3}. These PMUs can not be detected by Taobao, only being reported by retailers in the website (www.taocece.com):
	
	\begin{figure}[htbp]
		\centering
		\includegraphics[width=0.8\columnwidth]{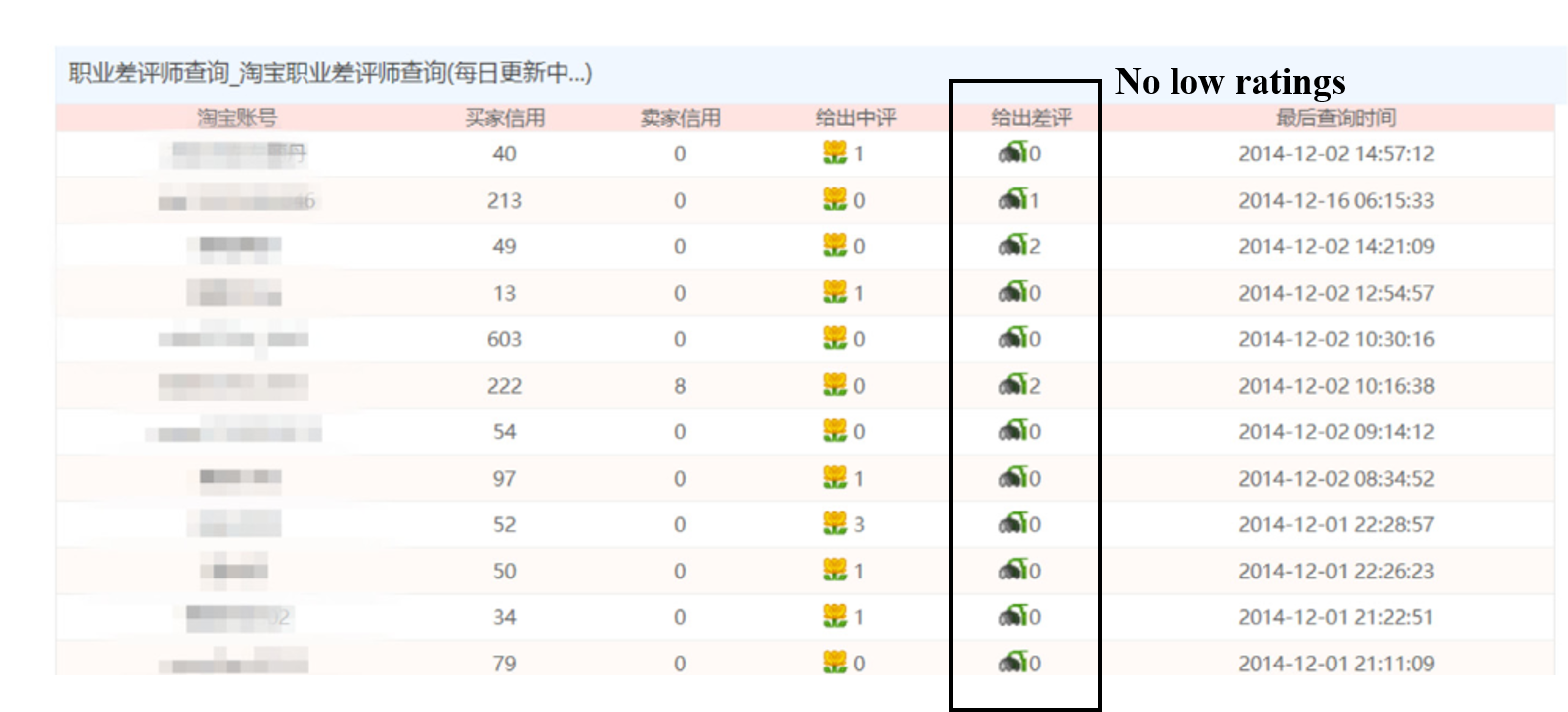}
		\caption{A real world case in Taobao.}
		\label{MMD4-3}
	\end{figure}
	
	Note that these PMUs do not usually give low ratings, so they can not be detected by traditional malicious user detection models. However, they usually give negative reviews with high ratings, which hurts retailers' profits and forms sentiment gaps between ratings and reviews. Our proposed model can catch the gap and detect this kind of PMUs.
	\subsection{Effect of Metric Learning (RQ2)}
	In this section, we explore the effect of metric learning in our model MMD, which consists of two parts: metric analysis and attention analysis.
	
	\subsubsection{Metric Analysis}
\begin{figure}[tbp]
		\centering
		\includegraphics[width=0.6\columnwidth]{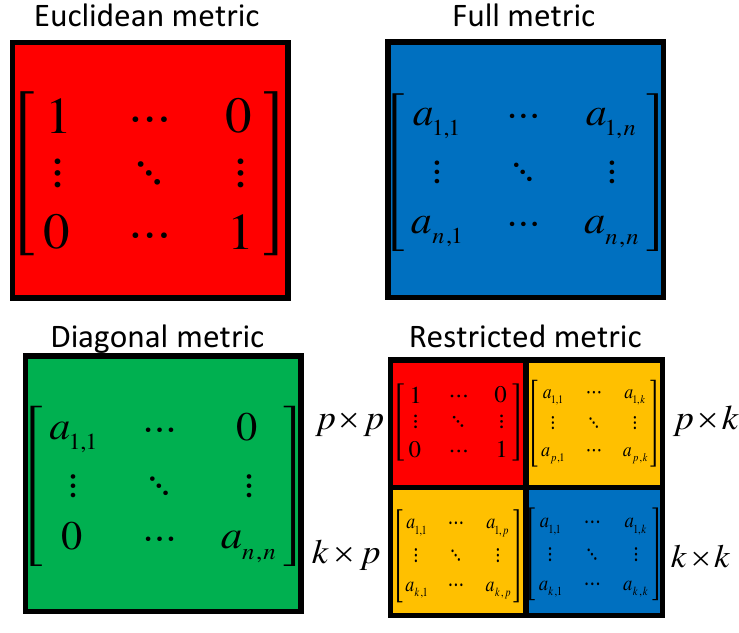}
		\caption{An illustration of different metrics}
		\label{MMD8}
	\end{figure}

	\begin{table*}[h]
		\centering
		\caption{Detection performance with different metrics. \textbf{bold} stands for MMD and $^*$ marks the best performance.}
		\label{MDtable7}
		\resizebox{0.9\textwidth}{!}{
			\begin{tabular}{llllllllllllll}
				\hline\hline
				&    \multicolumn{3}{c}{Amazon} & \multicolumn{3}{c}{Yelp}& \multicolumn{3}{c}{Jingdong}& \multicolumn{3}{c}{Jingdong}
				\\ \hline
				& \textit{SEN}  & \textit{SPE}  &F-score& \textit{SEN}  & \textit{SPE}  &F-score& \textit{SEN}  & \textit{SPE}  &F-score & \textit{SEN}  & \textit{SPE}  &F-score
				\\ \hline
				E-MMD&0.183&0.261&0.215&0.113&0.137&0.123&0.143&0.222&0.173&0.133&0.141&0.136
				\\ \hline
				D-MMD&0.395&0.662&0.494&0.547&0.633&0.587&0.657&0.642&0.649&0.589&0.492&0.536
				\\ \hline
				F-MMD&0.902&0.93&0.914&0.96&0.977&0.968&0.945$^*$&0.976&0.960&0.951&0.939&0.945
				\\ \hline
				R-MMD&\textbf{0.92$^*$}&\textbf{0.97$^*$}&\textbf{0.944$^*$}&\textbf{0.98$^*$}&\textbf{0.996$^*$}&\textbf{0.987$^*$}&\textbf{0.94}&\textbf{0.987$^*$}&\textbf{0.963$^*$}&\textbf{0.96$^*$}&\textbf{0.989$^*$}&\textbf{0.974$^*$}
				\\ \hline
		\end{tabular}}
	\end{table*}
	To verify the effect of metric, we apply different metric matrix A on MLC: Euclidean metric MMD (E-MMD), Diagonal-matrix MMD (D-MMD), and Full-matrix MMD (F-MMD) and our restricted matrix metric MMD (R-MMD). Specifically, E-MMD aims to learn the Euclidean metric, where it is a scalar; D-MMD aims to learn a diagonal matrix, which assigns a weight vector for Euclidean metric; F-MMD is to gain a full matrix, which is a general metric learning method. Our R-MMD is to learn a p-dimension diagonal matrix and a k-dimension full matrix, and two p$\times$k matrices. All these metrics are shown in Fig. \ref{MMD8}.
	
	We show the detection performance with different metrics in Table \ref{MDtable7}. We notice that among all the datasets, E-MMD performs worst, and our R-MMD performs best. Specifically, we validate that Euclidean distance (E-MMD) can not measure the relationship between users' attributes and sentiment gaps, also the weighted Euclidean (D-MMD). However, if we set the metric matrix to a full matrix, it can achieve the same performance as our R-MMD. However, the number of F-MMD parameters is much more substantial than R-MMD (in our experiment, F-MMD needs 4096 parameters, while R-MMD needs 3104 parameters), which may cost more computing resources. 
	
	\subsubsection{Attention Analysis}
	To explore the attention mechanism for our method, we conduct MMD on all four datasets without applying attention as a reference group. We show the results in Table \ref{MDtable8}. Note that the model with attention mechanism outperforms the no-attention model on all datasets and improves the detection performance by average 7.67\%. The attention mechanism is proper to catch the dynamic relationships, which is suitable for professional malicious user detections. Note that in our method, the attention vector can indicate the relationships not only between attributes and attributes, sentiment gap and sentiment gap, but the relationships across them, which offers the probability for explainability.
	\begin{table*}[h]
		\centering
		\caption{Effect of attention mechanism on MMD for malicious detections. \textbf{bold} stands for MMD and $^*$ marks the best performance.}
		\label{MDtable8}
		\resizebox{0.99\textwidth}{!}{
			\begin{tabular}{llllllllllllll}
				\hline\hline
				&    \multicolumn{3}{c}{Amazon} & \multicolumn{3}{c}{Yelp}& \multicolumn{3}{c}{Jingdong}& \multicolumn{3}{c}{Jingdong}
				\\ \hline
				MMD& \textit{SEN}  & \textit{SPE}  &F-score& \textit{SEN}  & \textit{SPE}  &F-score& \textit{SEN}  & \textit{SPE}  &F-score & \textit{SEN}  & \textit{SPE}  &F-score
				\\ \hline
				\textit{noatt}&0.87&0.912&0.890&0.884&0.914&0.899&0.834&0.913&0.872&0.924&0.939&0.931
				\\ \hline
				\textit{att}&\textbf{0.92$^*$}&\textbf{0.97$^*$}&\textbf{0.944$^*$(+6.0\%)}&\textbf{0.98$^*$}&\textbf{0.996$^*$}&\textbf{0.987$^*$(+9.7\%)}&\textbf{0.94$^*$}&\textbf{0.987$^*$}&\textbf{0.963$^*$+(10.4\%)}&\textbf{0.96$^*$}&\textbf{0.989$^*$}&\textbf{0.974$^*$(+4.6\%)}
				\\ \hline
		\end{tabular}}
	\end{table*}
	\subsection{Parameter Exploration (RQ3)}
	In this subsection, we examine the impacts of parameters, i.e., $\alpha$ and $\theta$, which control the sentiment gap to profile professional malicious users and the scale of the reviews with the sentiment gap of each user. When we explore the effect of the changing parameter, all other parameters are fixed to the initialized values.
	\begin{figure}[!h] \centering  
		\subfigure[F-score on Amazon] { \label{fig:subfig:a1}     
			\includegraphics[width=0.46\columnwidth]{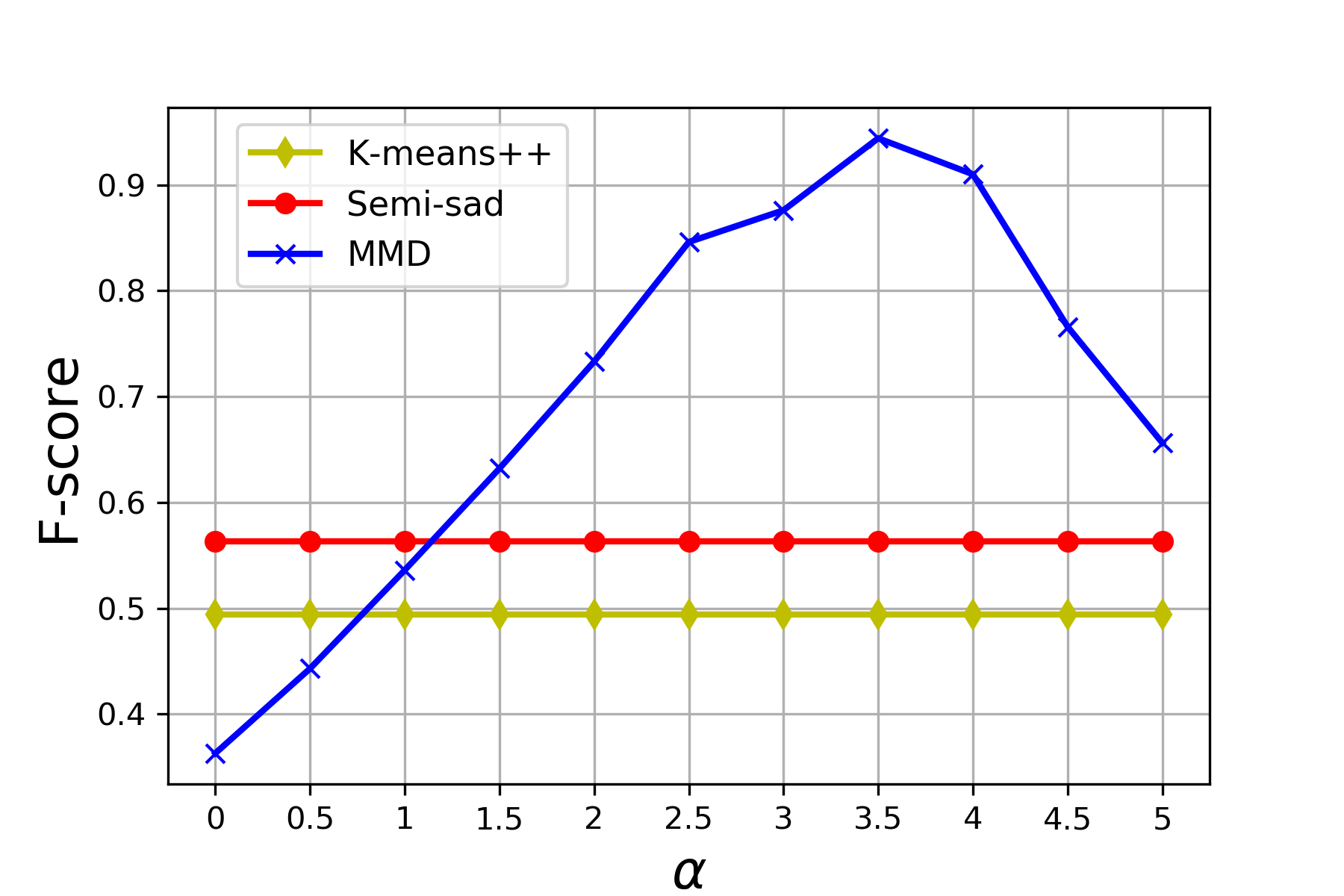} 
			\vspace{-10pt}  
		}     
		\subfigure[F-score on Yelp] { \label{fig:subfig:a2}     
			\includegraphics[width=0.46\columnwidth]{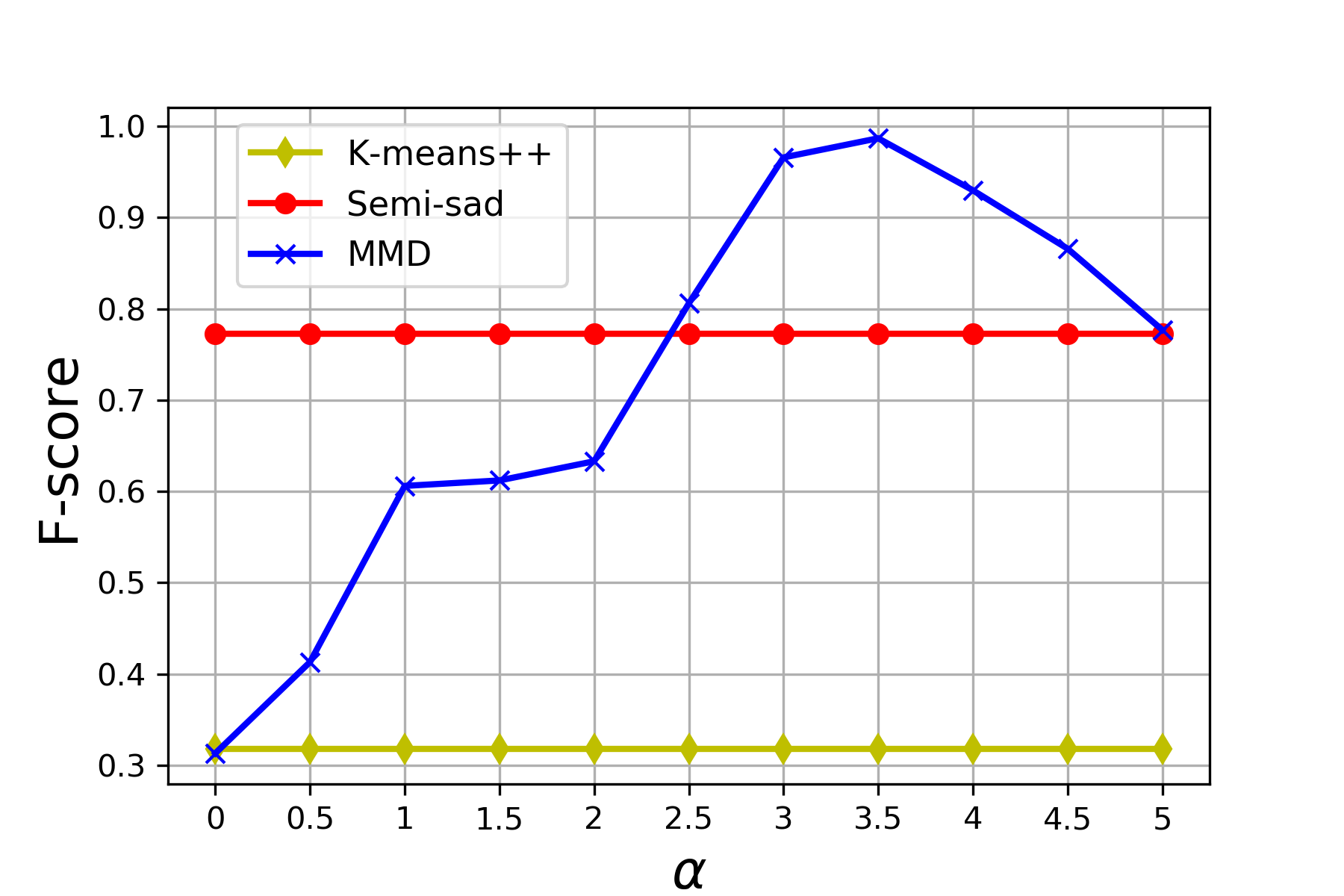} 
			\vspace{-10pt}     
		}
		\subfigure[F-score on Taobao] { \label{fig:subfig:a3}     
			\includegraphics[width=0.46\columnwidth]{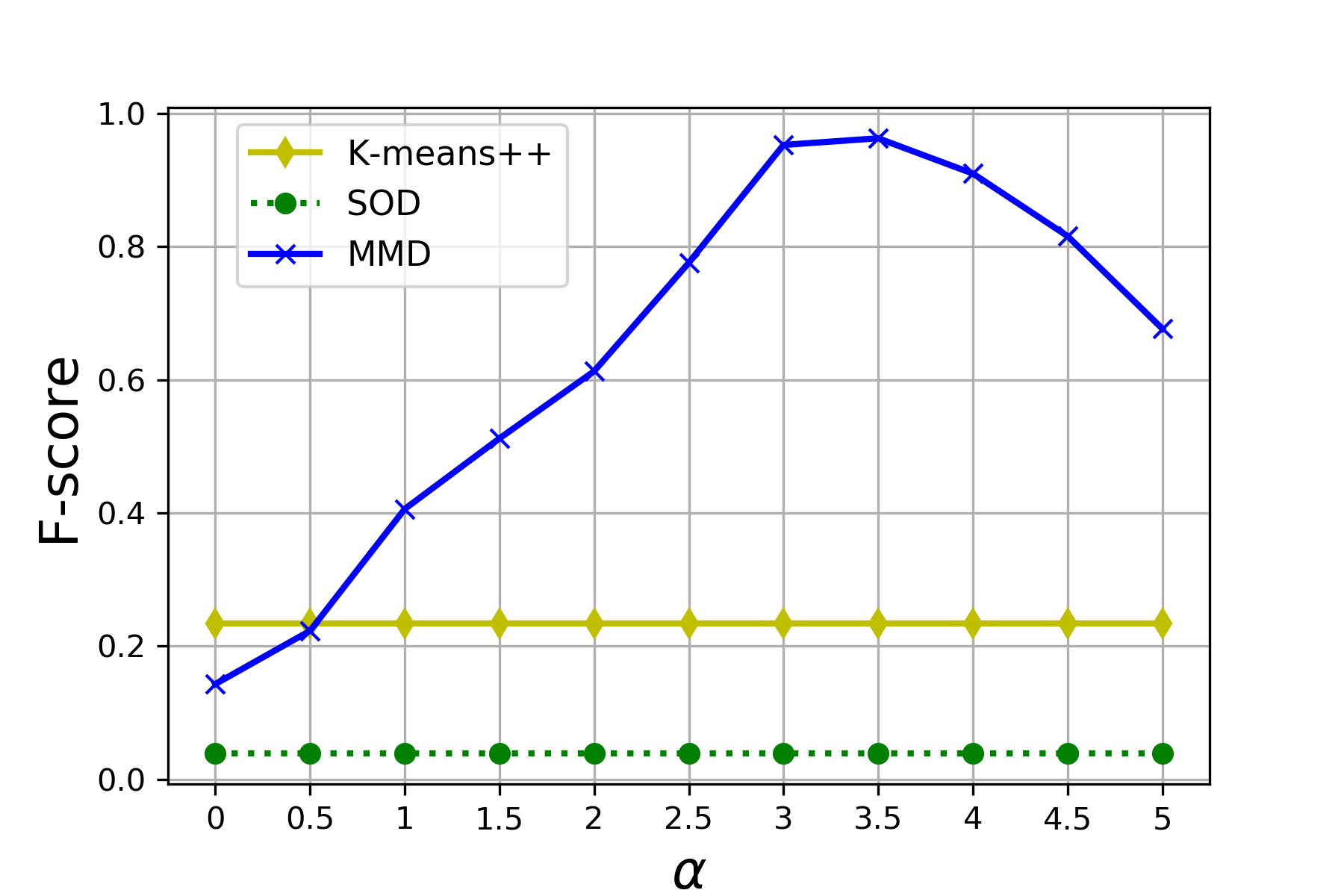} 
			\vspace{-10pt}  
		}     
		\subfigure[F-score on Jingdong] { \label{fig:subfig:a4}     
			\includegraphics[width=0.46\columnwidth]{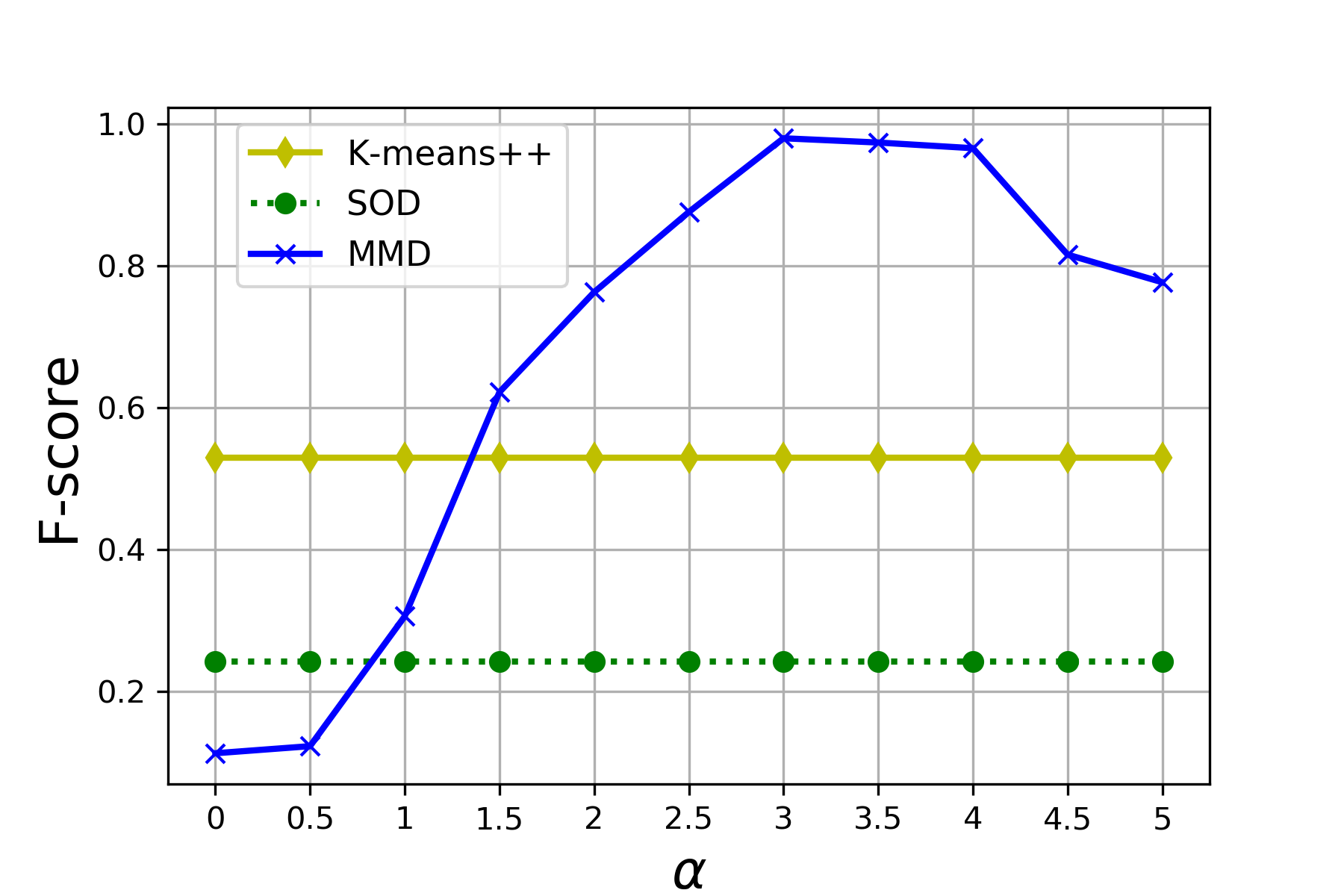} 
			\vspace{-10pt}     
		}       
		\caption{Performance of MMD \textit{w.r.t.} different values of $\alpha$. MMD achieves the best F-score performance when $\alpha$=3.5, 3.5, 3.5 and 3 on Amazon, Yelp, Taobao and Jingdong, respectively.}     
		\label{MMD4}     
	\end{figure}
	\begin{figure}[!h] \centering  
		\subfigure[F-score on Amazon] { \label{fig:subfig:b1}     
			\includegraphics[width=0.46\columnwidth]{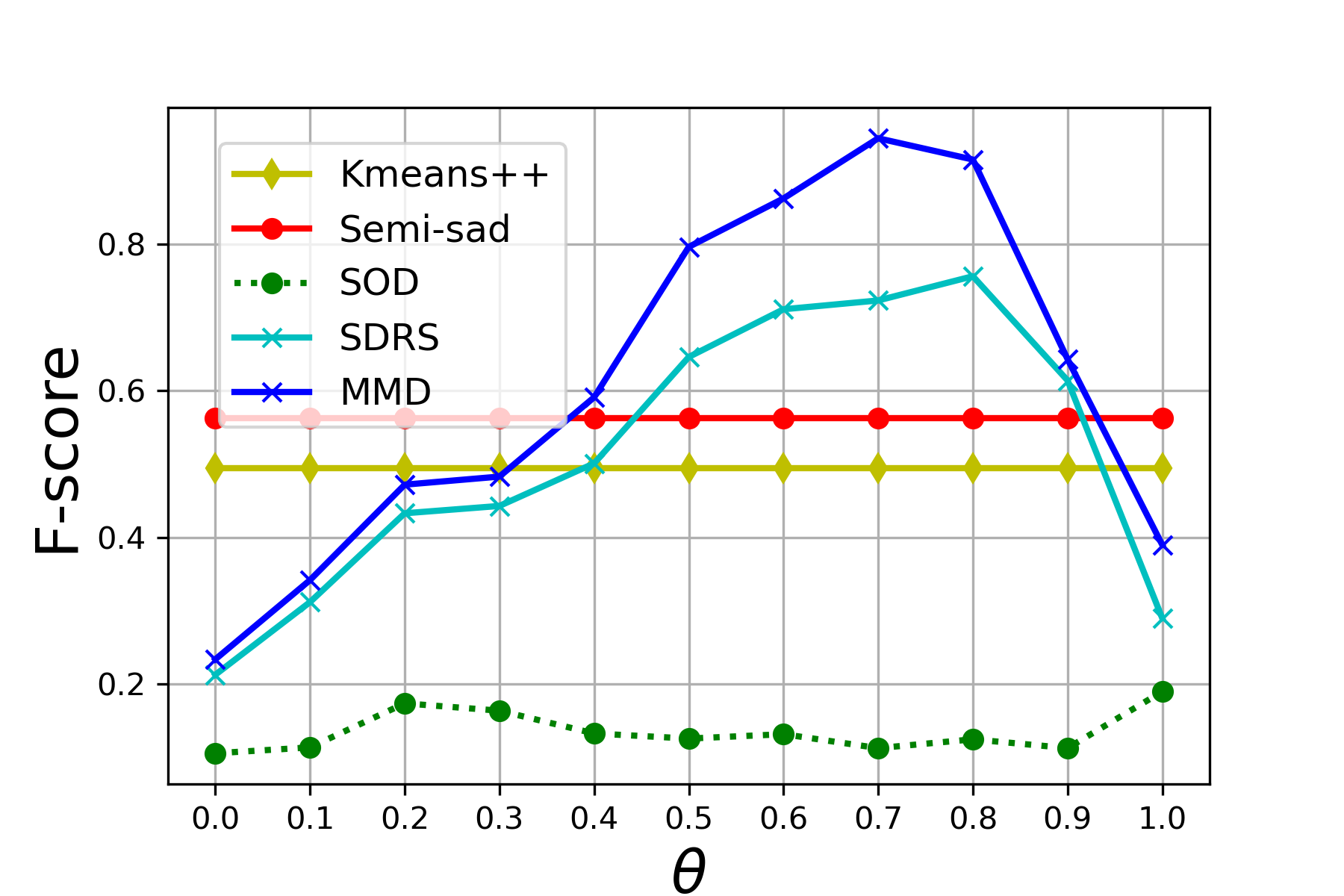} 
			\vspace{-10pt}  
		}     
		\subfigure[F-score on Yelp] { \label{fig:subfig:b2}     
			\includegraphics[width=0.46\columnwidth]{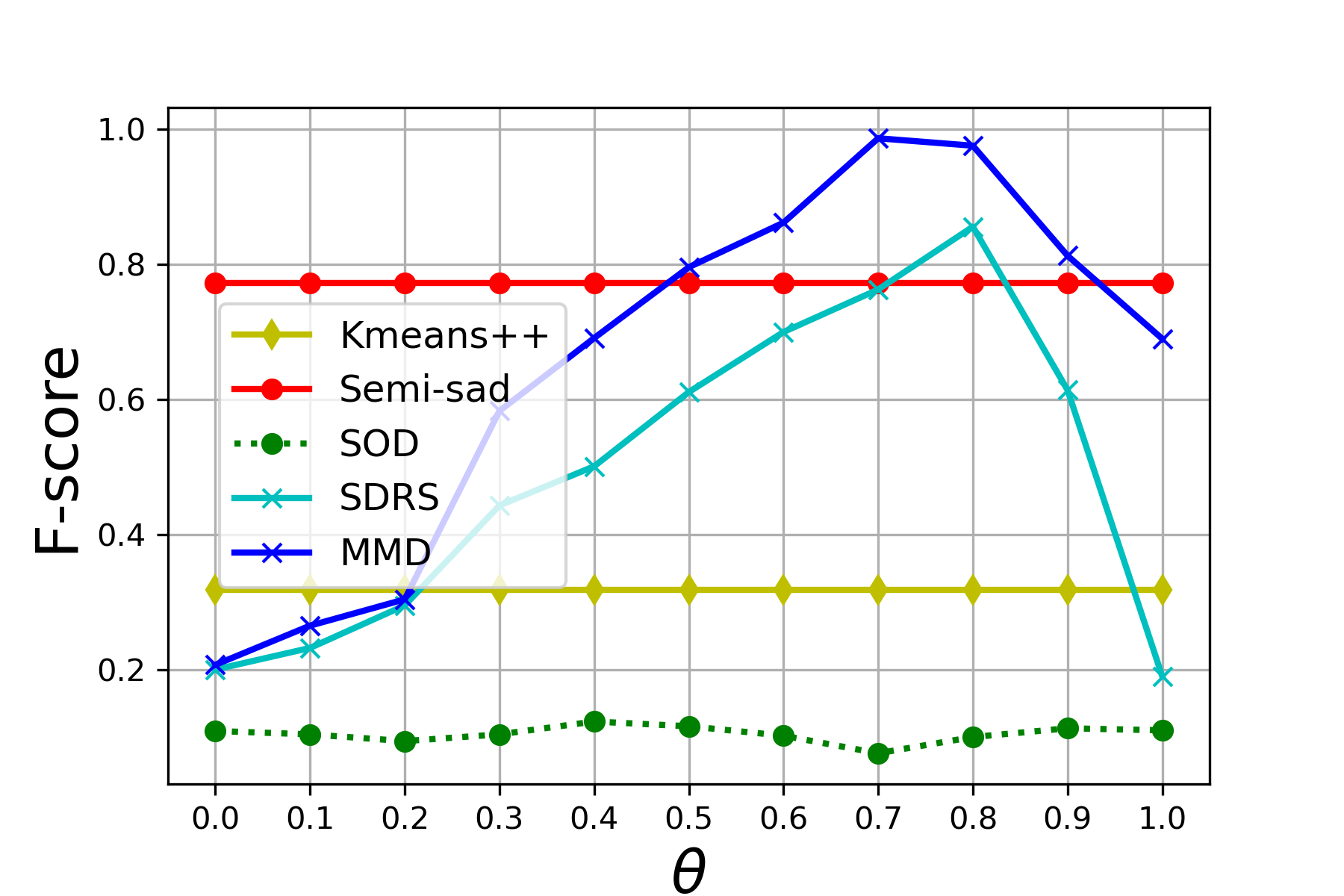} 
			\vspace{-10pt}     
		}
		\subfigure[F-score on Taobao] { \label{fig:subfig:b3}     
			\includegraphics[width=0.46\columnwidth]{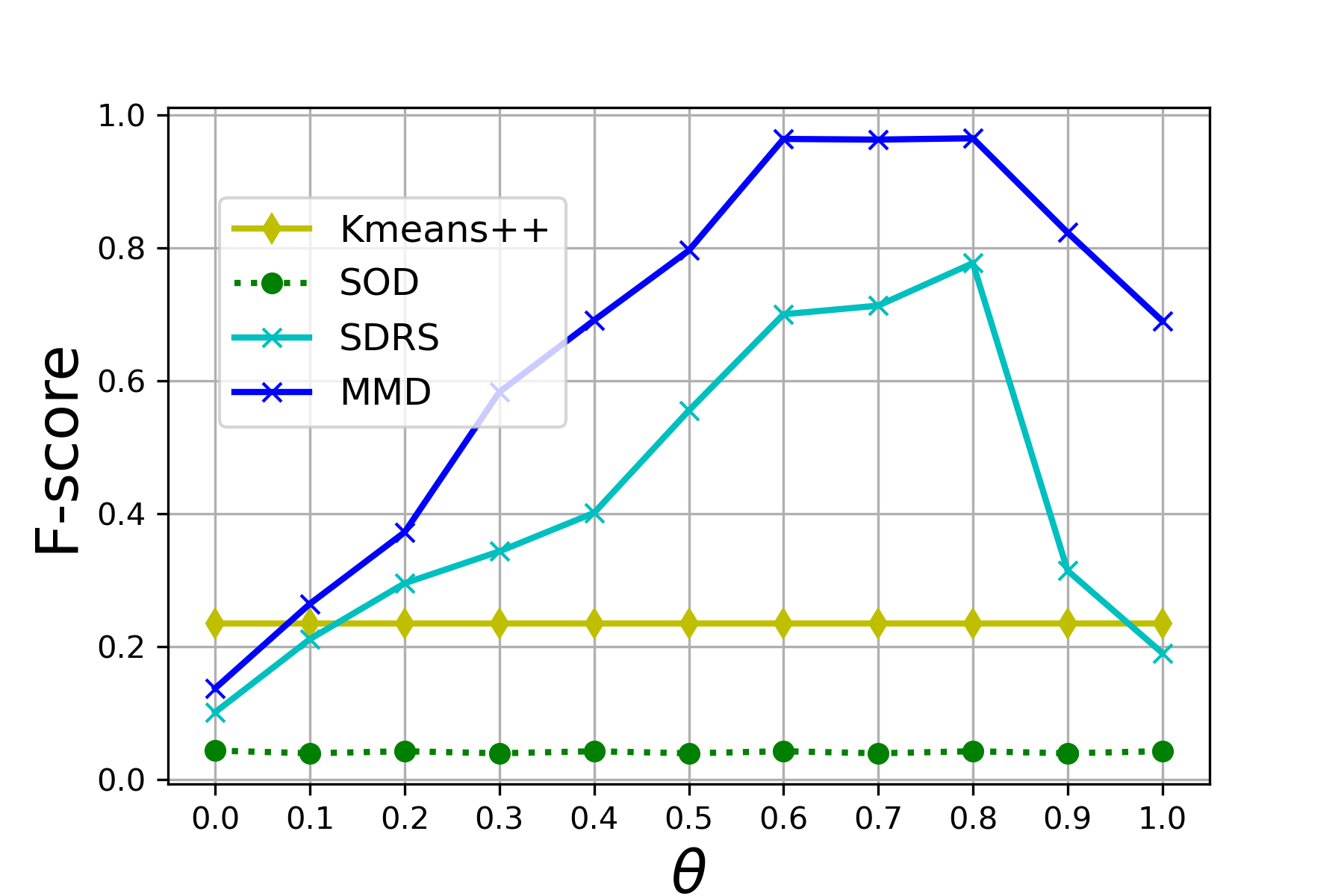} 
			\vspace{-10pt}  
		}     
		\subfigure[F-score on Jingdong] { \label{fig:subfig:b4}     
			\includegraphics[width=0.46\columnwidth]{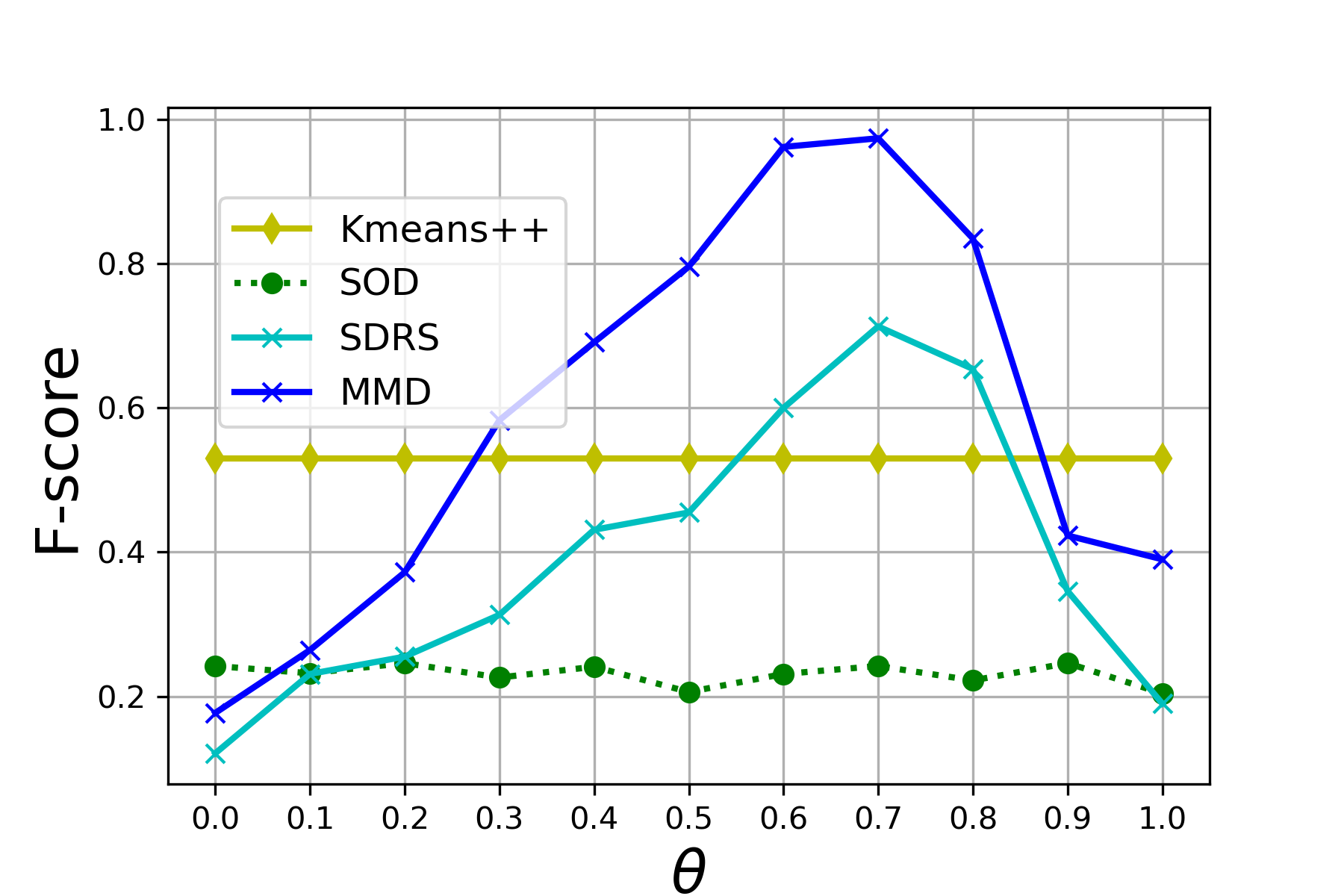} 
			\vspace{-10pt}     
		}       
		\caption{Performance of MMD \textit{w.r.t.} different values of $\theta$. MMD achieves the best F-score performance when $\theta$=0.7, 0.7, 0.8 and 0.7 on Amazon, Yelp, Taobao and Jingdong, respectively.}     
		\label{MMD5}     
	\end{figure}
	
	Fig.\ref{MMD4} illustrates the performance changes with respect to $\alpha$. Note that our proposed model MMD achieves the optimal F-score performance when $\alpha$=3.5, 3.5, 3.5, and 3 on Amazon, Yelp, Taobao, and Jingdong, respectively. So we set $\alpha$=3.5 as default. When $\alpha$ is smaller than 2, increasing it leads to gradual improvement. In detail, we notice that SEN is low in this situation, which means that the model can not detect professional malicious users correctly. The result implies that the sentiment gap is always larger than 2 for PMUs, and the gap exists in normal users when it is smaller than 2. When $\alpha$ is larger than the optimal point (3 or 3.5), the performance drops rapidly, which means that the bigger gap threshold will affect the SPE and lead to bad detections. 
	
	Fig.\ref{MMD5} illustrates the performance with respect to $\theta$. Note that our proposed model MMD achieves the optimal F-score performance when $\alpha$=0.7, 0.7, 0.8, and 0.7 on Amazon, Yelp, Taobao, and Jingdong, respectively. So we set $\theta$=0.7 as default. When $\theta$ is smaller than the threshold, the F-score improves gradually. When it adds over 0.6, MMD achieves a relatively stable performance (much better than baselines among all the datasets). Note that when $\theta$ is close to 1, it means that the model wants to detect the users with all fake feedbacks. In this situation, F-score drops rapidly because the model loses the ability to find professional malicious users. The professional malicious users do not give all fake feedbacks, which also proves our definition of professional malicious users. Specifically, the optimal points of the parameters are different for the four datasets, which indicates that for different datasets, the parameters should be separately tuned to achieve the best performance.\begin{figure*}[!h] \centering  
		\subfigure[HR@5 on Amazon] { \label{fig:subfig:c1}     
			\includegraphics[width=0.47\columnwidth]{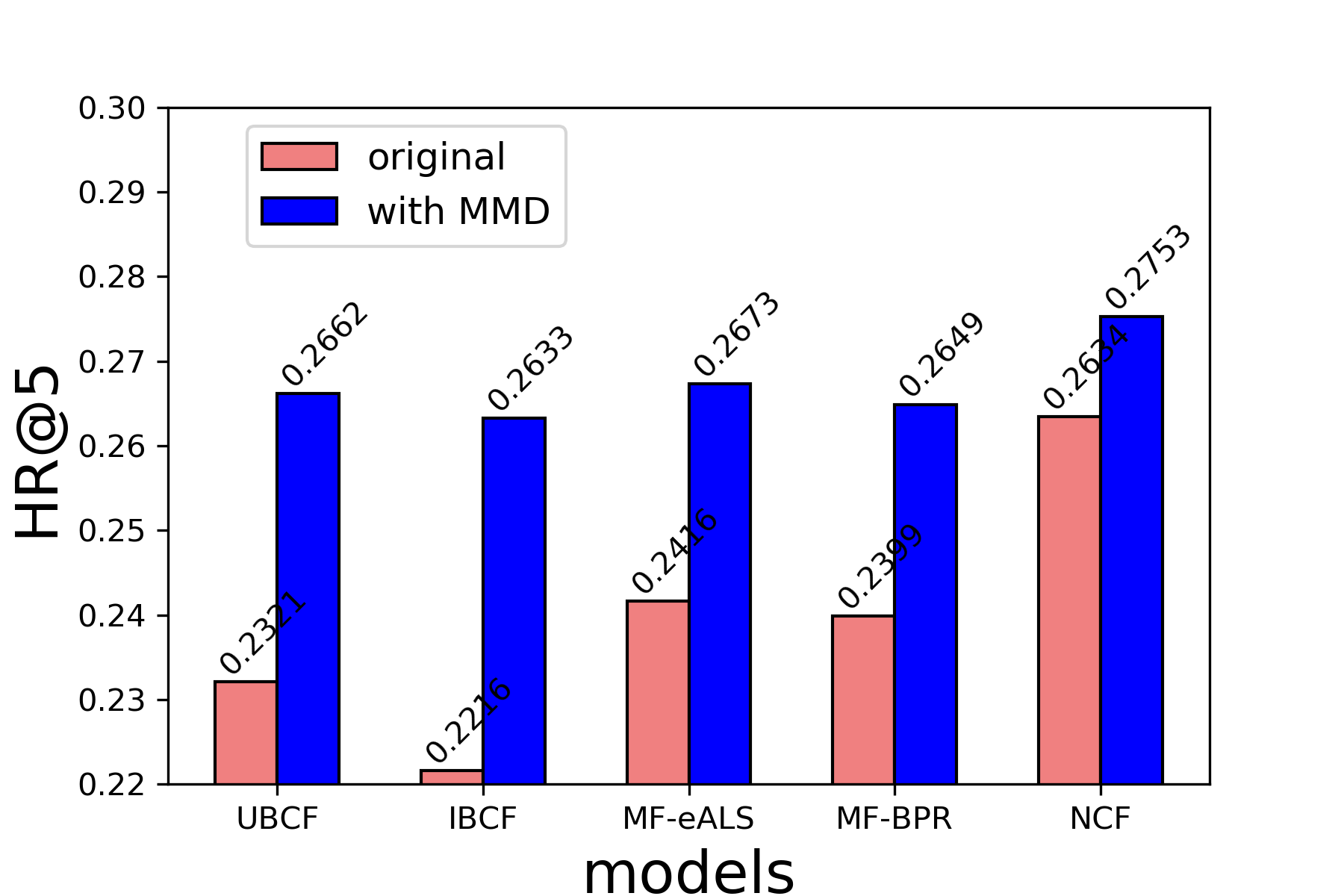} 
			\vspace{-10pt}  
		}     
		\subfigure[HR@5 on Yelp] { \label{fig:subfig:c2}     
			\includegraphics[width=0.47\columnwidth]{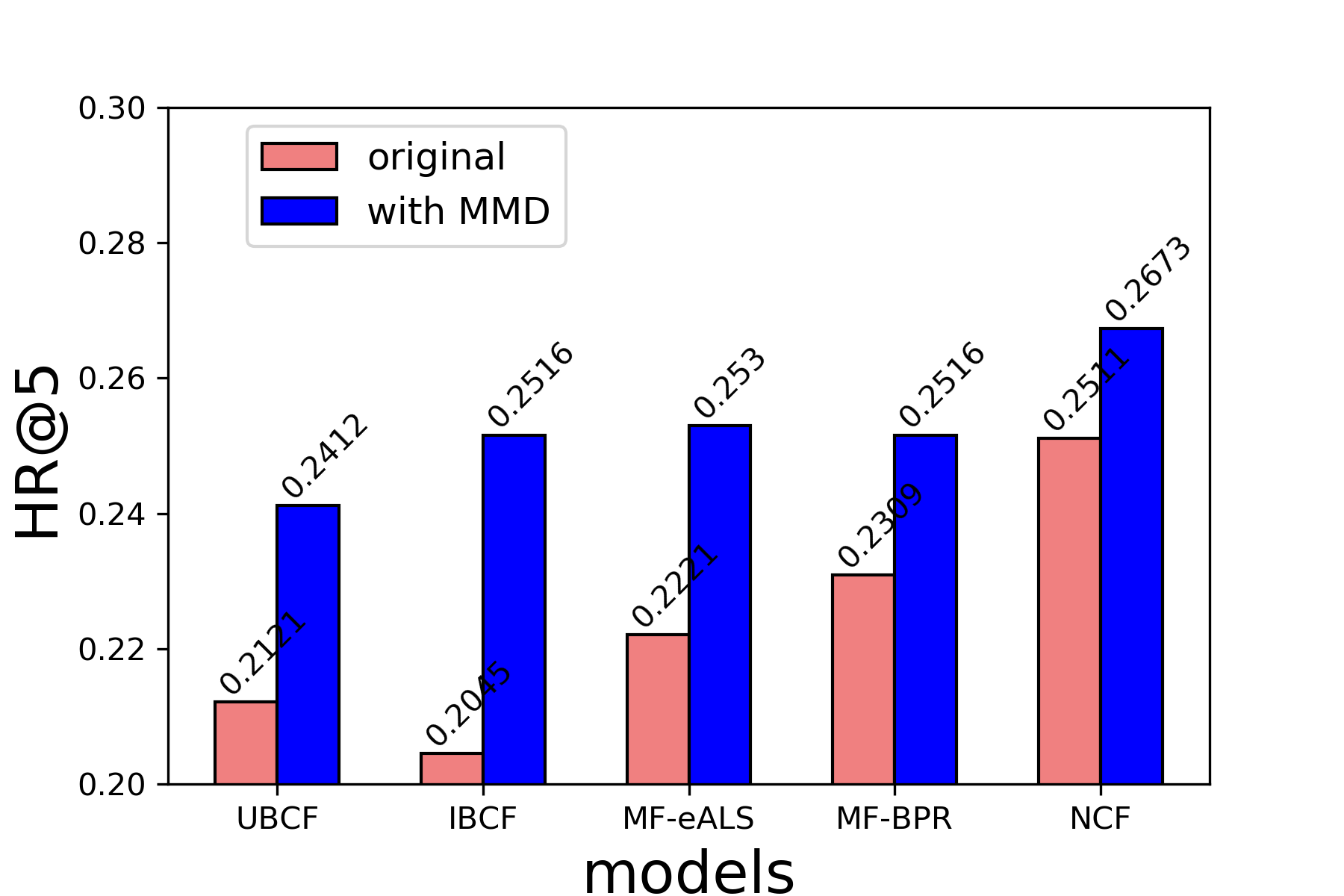} 
			\vspace{-10pt}     
		}
		\subfigure[HR@5 on Taobao] { \label{fig:subfig:c3}     
			\includegraphics[width=0.47\columnwidth]{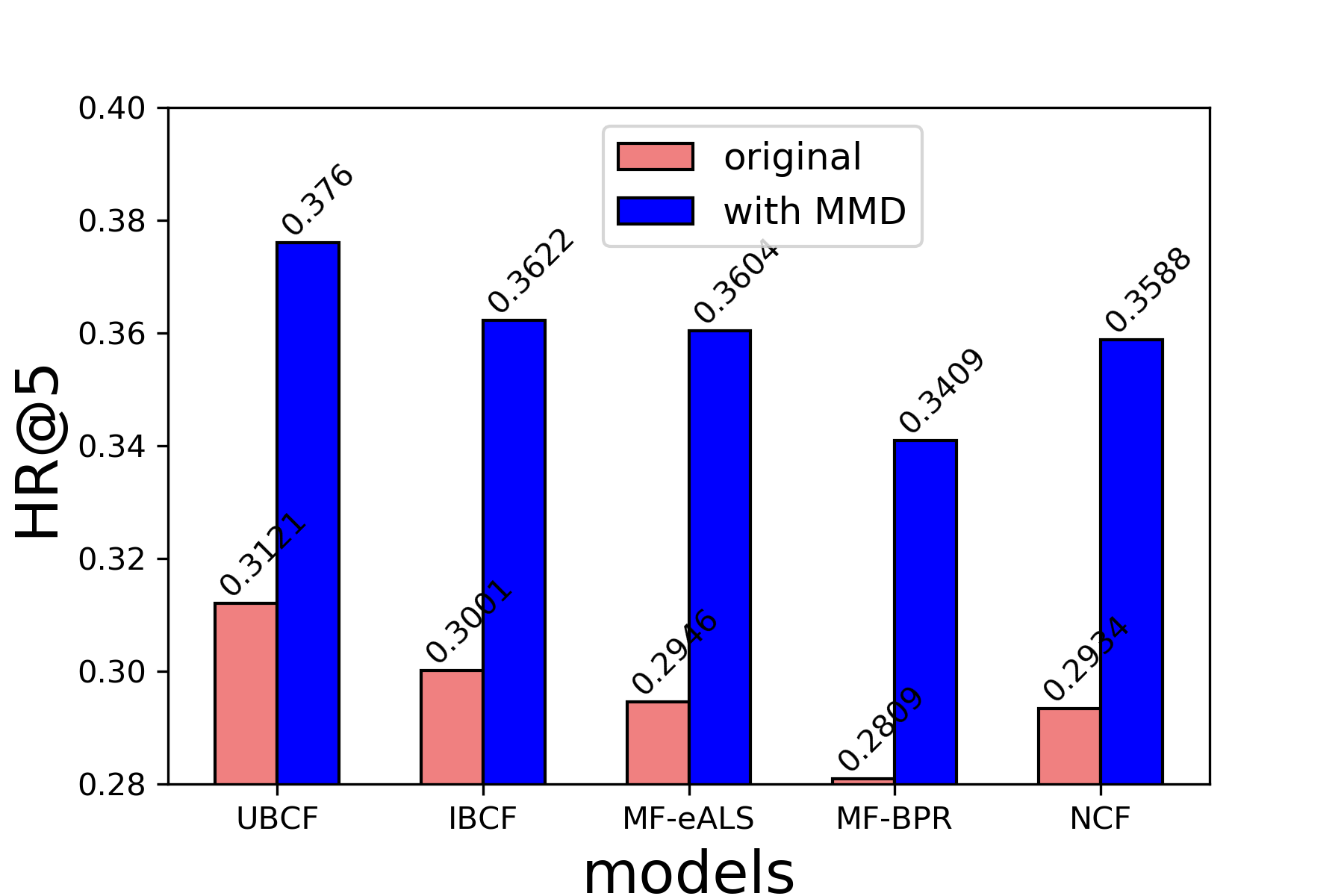} 
			\vspace{-10pt}  
		}     
		\subfigure[HR@5 on Jingdong] { \label{fig:subfig:c4}     
			\includegraphics[width=0.47\columnwidth]{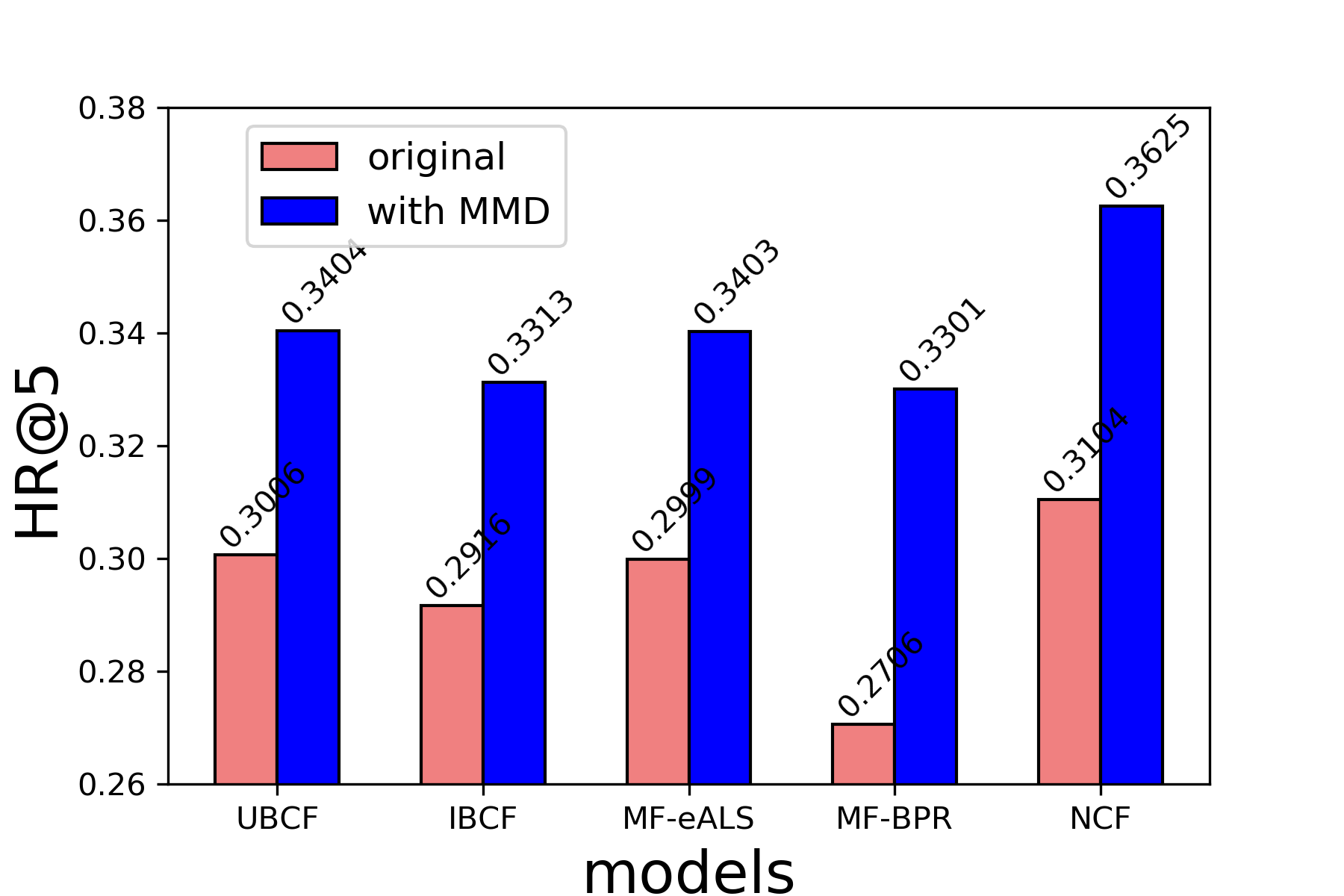} 
			\vspace{-10pt}     
		}       
		\subfigure[HR@15 on Amazon] { \label{fig:subfig:c5}     
			\includegraphics[width=0.47\columnwidth]{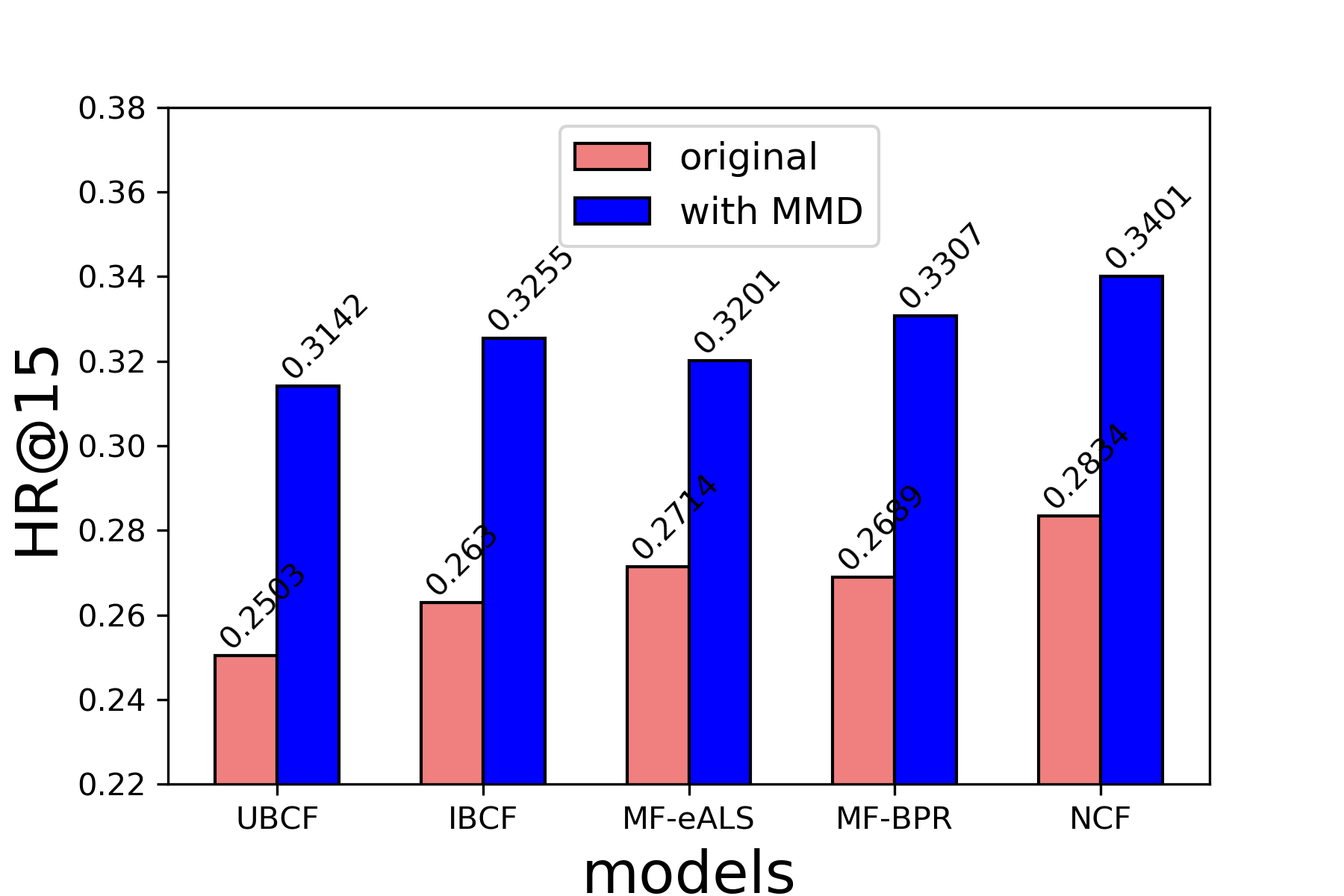} 
			\vspace{-10pt}  
		}     
		\subfigure[HR@15 on Yelp] { \label{fig:subfig:c6}     
			\includegraphics[width=0.47\columnwidth]{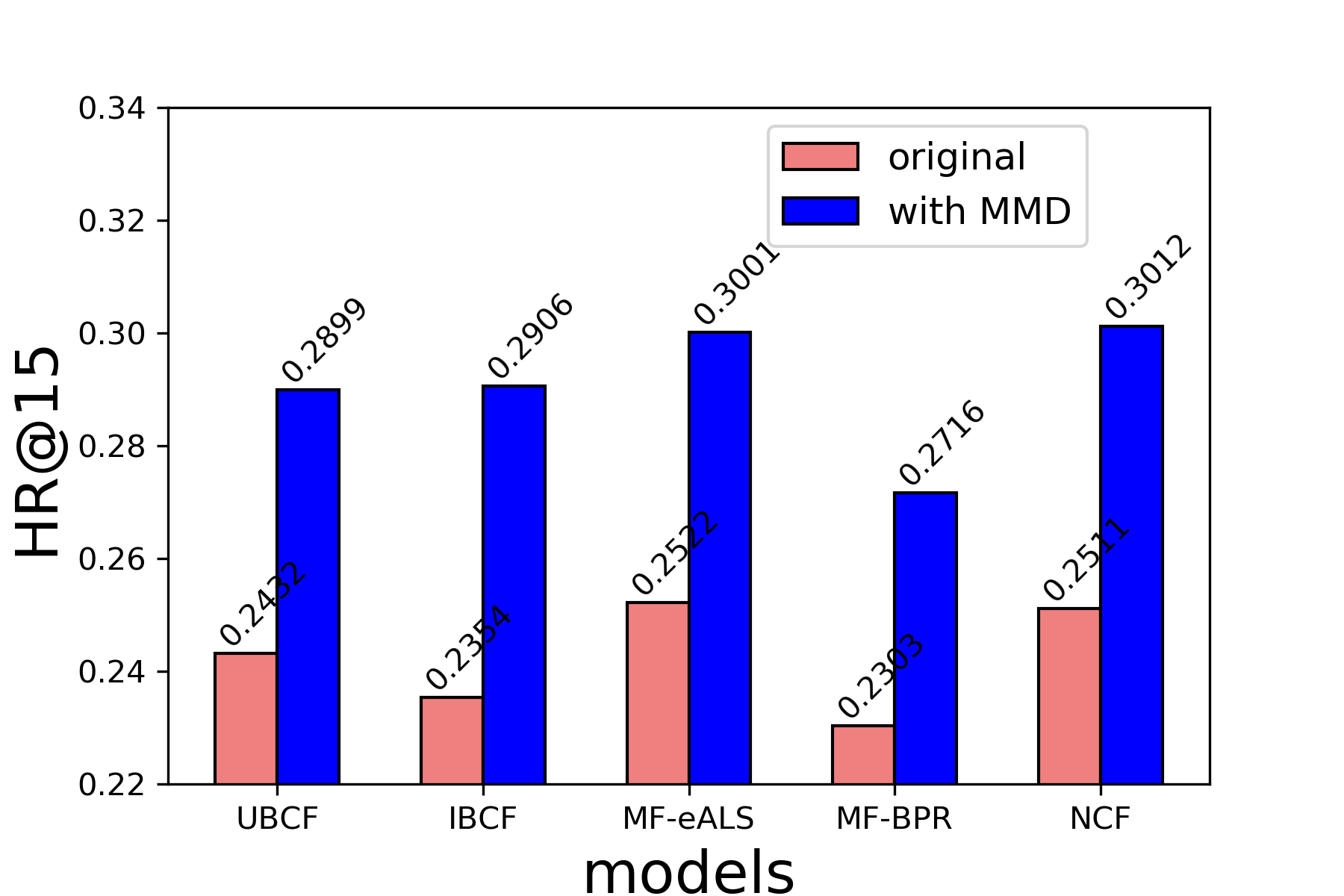} 
			\vspace{-10pt}     
		}
		\subfigure[HR@15 on Taobao] { \label{fig:subfig:c7}     
			\includegraphics[width=0.47\columnwidth]{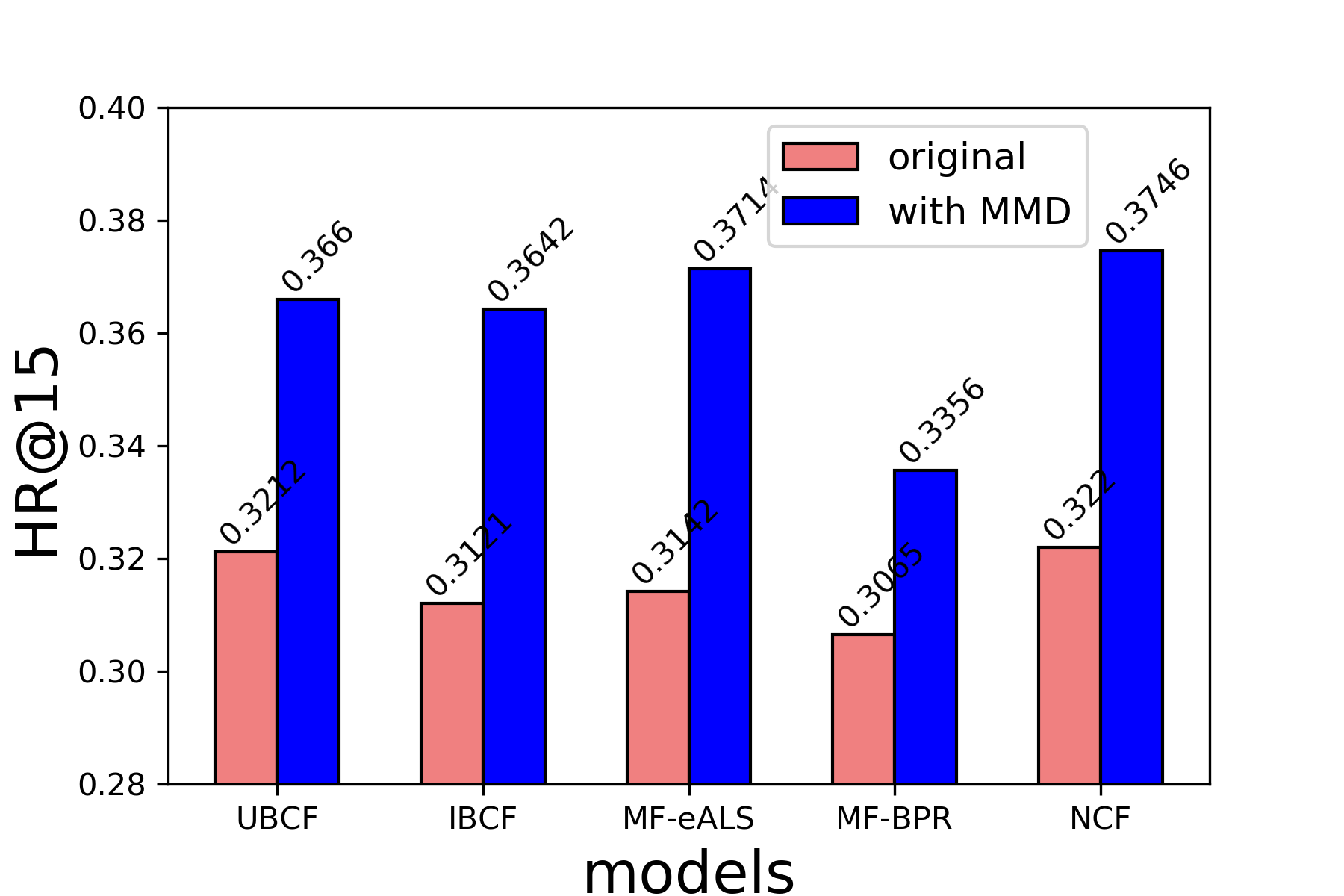} 
			\vspace{-10pt}  
		}     
		\subfigure[HR@15 on Jingdong] { \label{fig:subfig:c8}     
			\includegraphics[width=0.47\columnwidth]{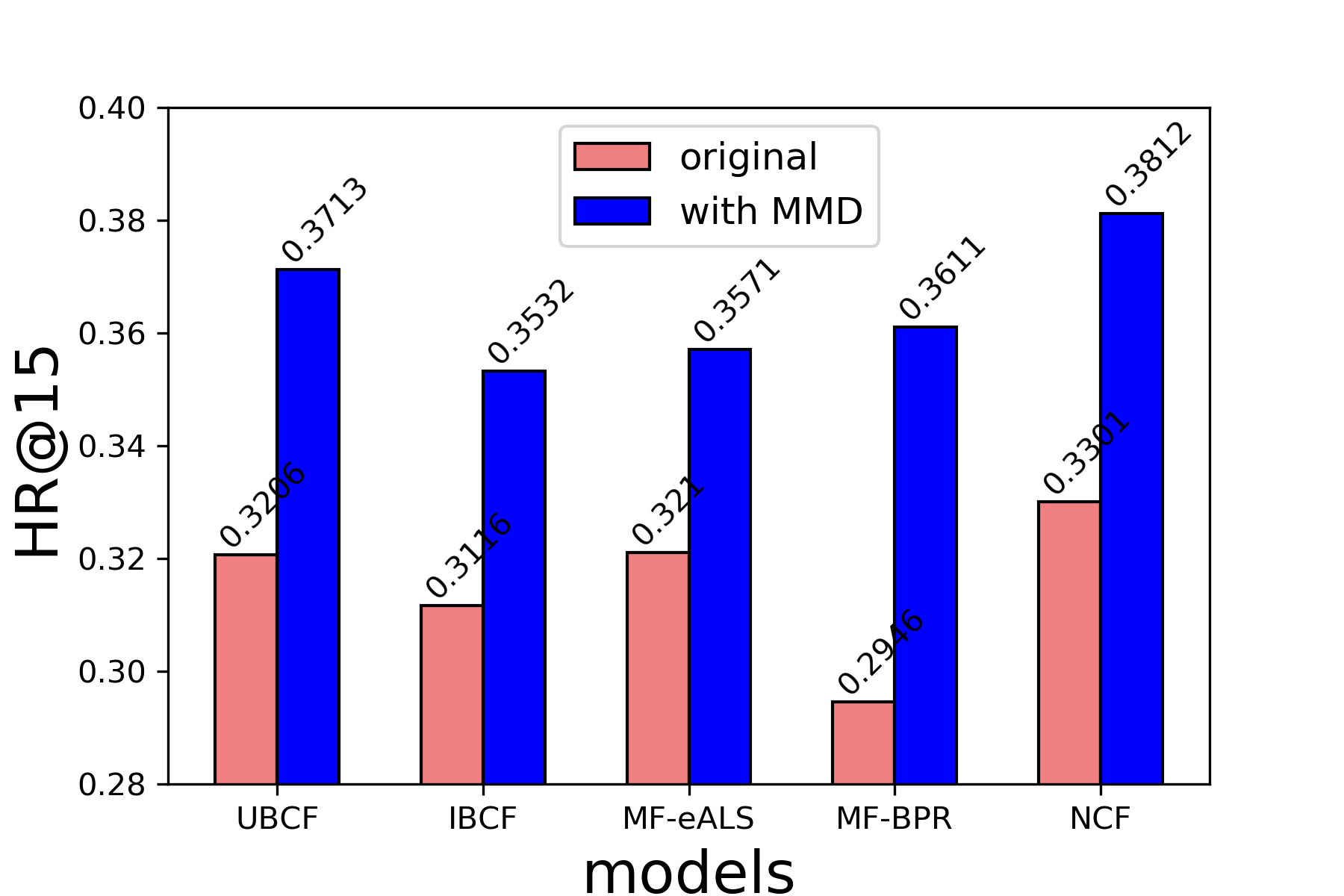} 
			\vspace{-10pt}     
		}   
		\caption{Top-N recommendation performance (HR) of different recommendation models with/without MMD as a preprocessing.}     
		\label{MMD6}     
	\end{figure*}
	\begin{figure*}[!h] \centering  
		\subfigure[NDCG@5 on Amazon] { \label{fig:subfig:d1}     
			\includegraphics[width=0.47\columnwidth]{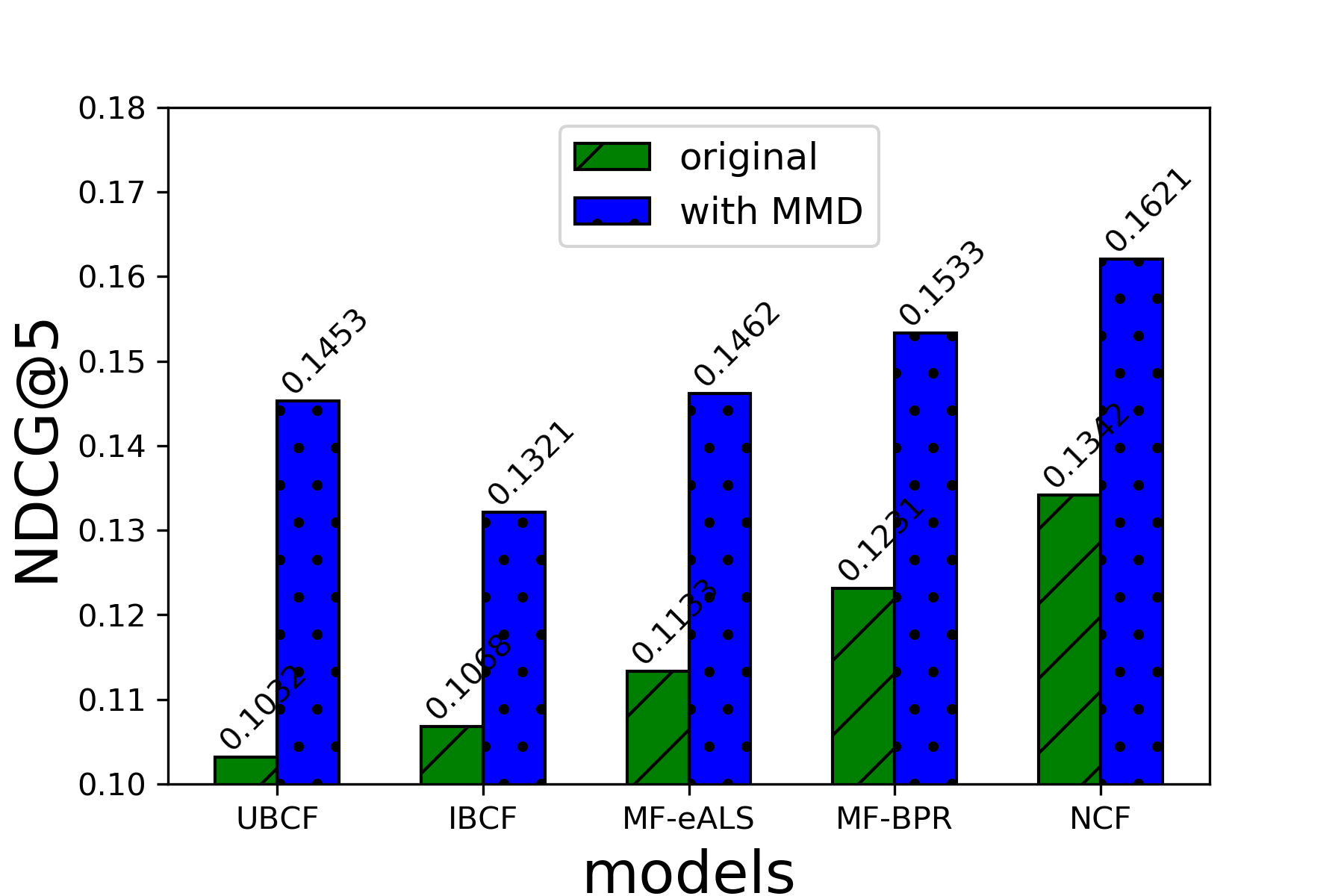} 
			\vspace{-10pt}  
		}     
		\subfigure[NDCG@5 on Yelp] { \label{fig:subfig:d2}     
			\includegraphics[width=0.47\columnwidth]{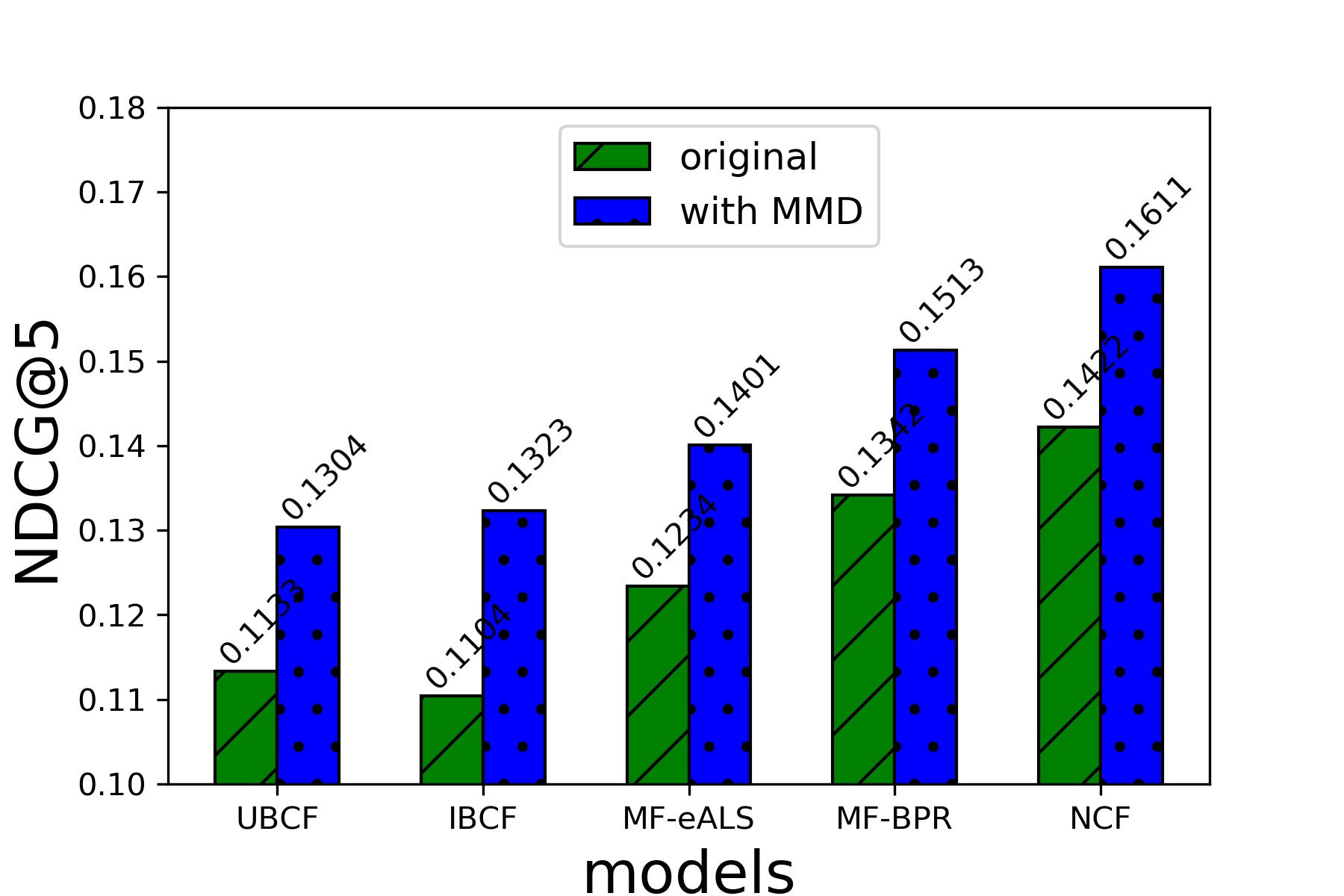} 
			\vspace{-10pt}     
		}
		\subfigure[NDCG@5 on Taobao] { \label{fig:subfig:d3}     
			\includegraphics[width=0.47\columnwidth]{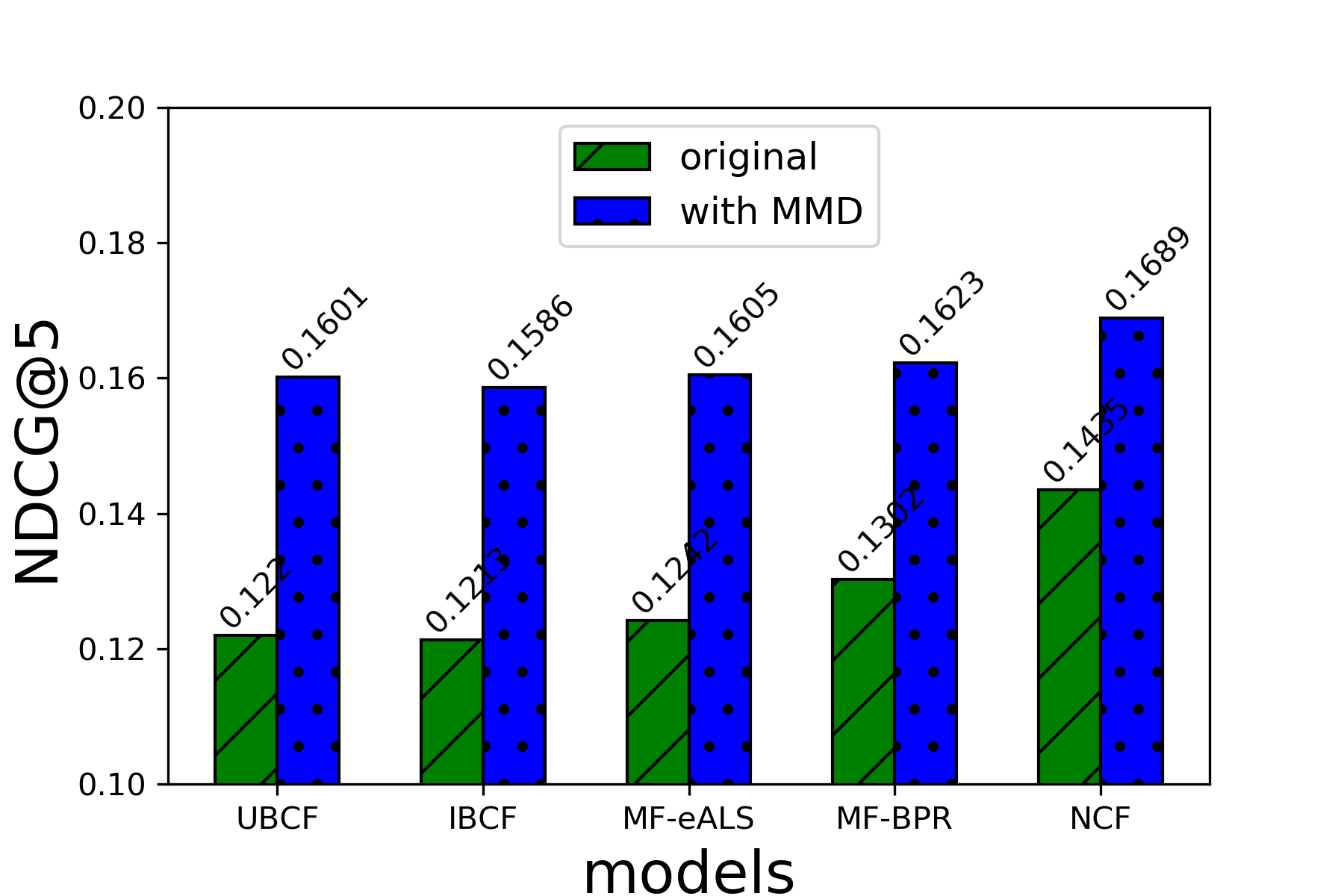} 
			\vspace{-10pt}  
		}     
		\subfigure[NDCG@5 on Jingdong] { \label{fig:subfig:d4}     
			\includegraphics[width=0.47\columnwidth]{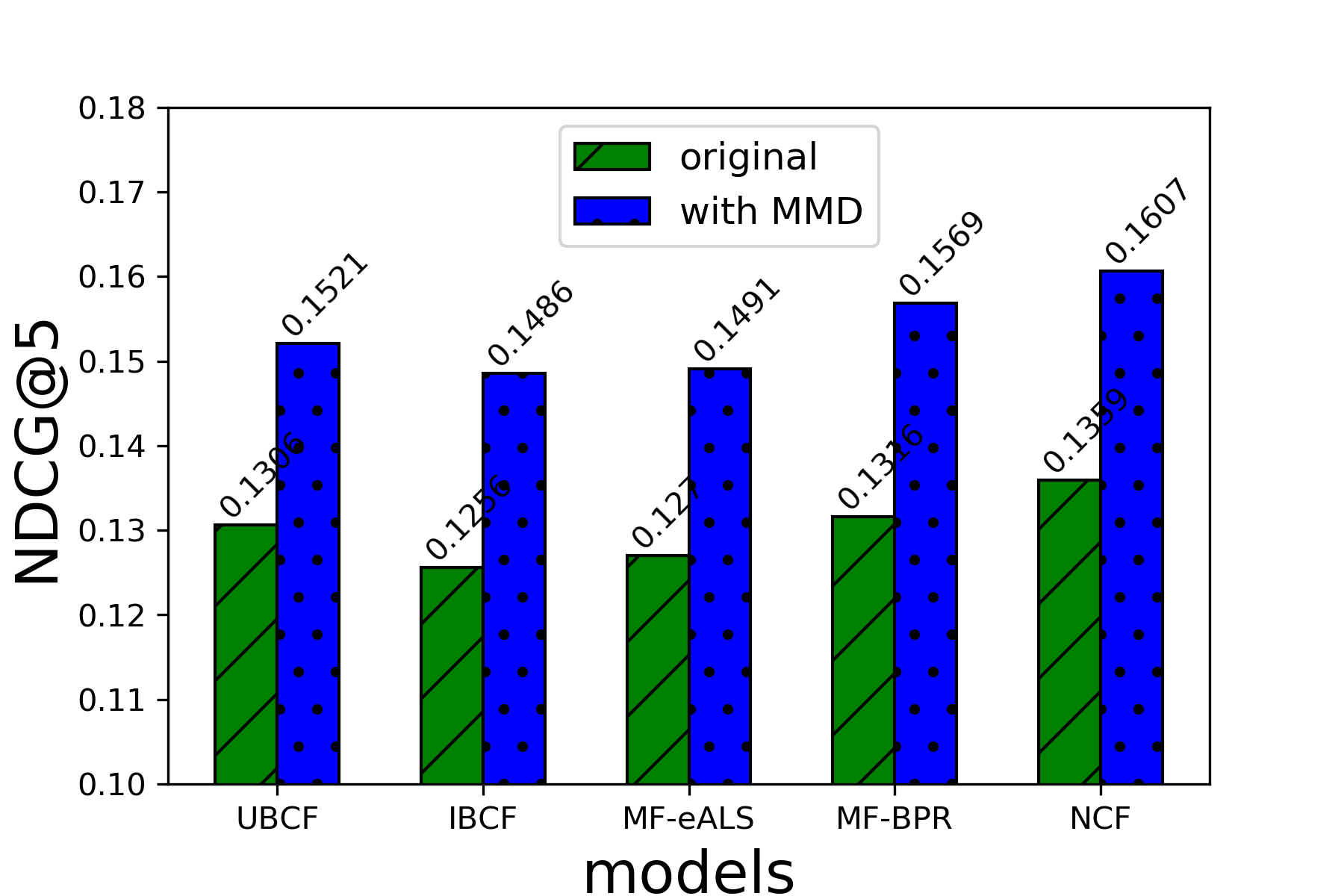} 
			\vspace{-10pt}     
		}       
		\subfigure[NDCG@15 on Amazon] { \label{fig:subfig:d5}     
			\includegraphics[width=0.47\columnwidth]{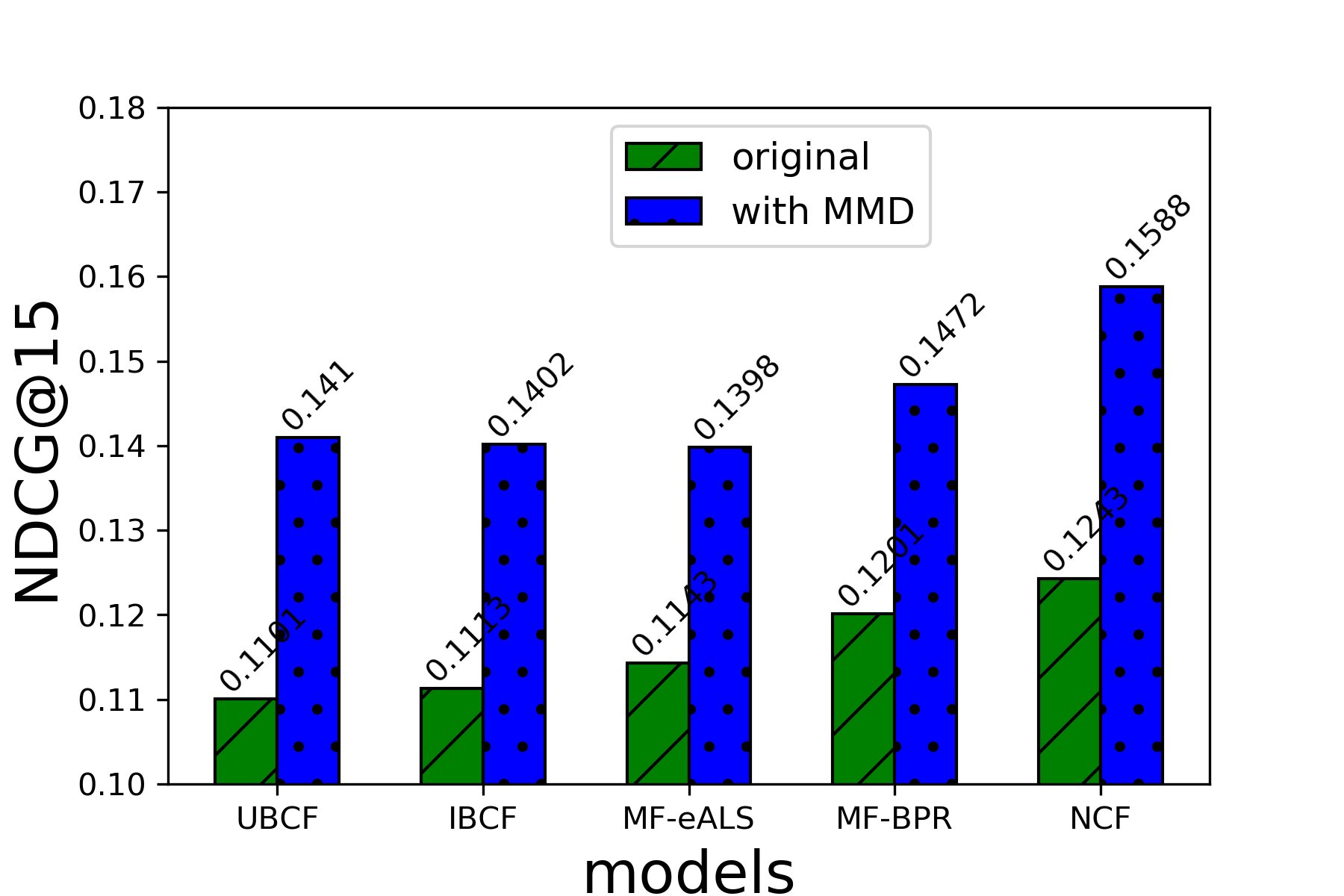} 
			\vspace{-10pt}  
		}     
		\subfigure[NDCG@15 on Yelp] { \label{fig:subfig:d6}     
			\includegraphics[width=0.47\columnwidth]{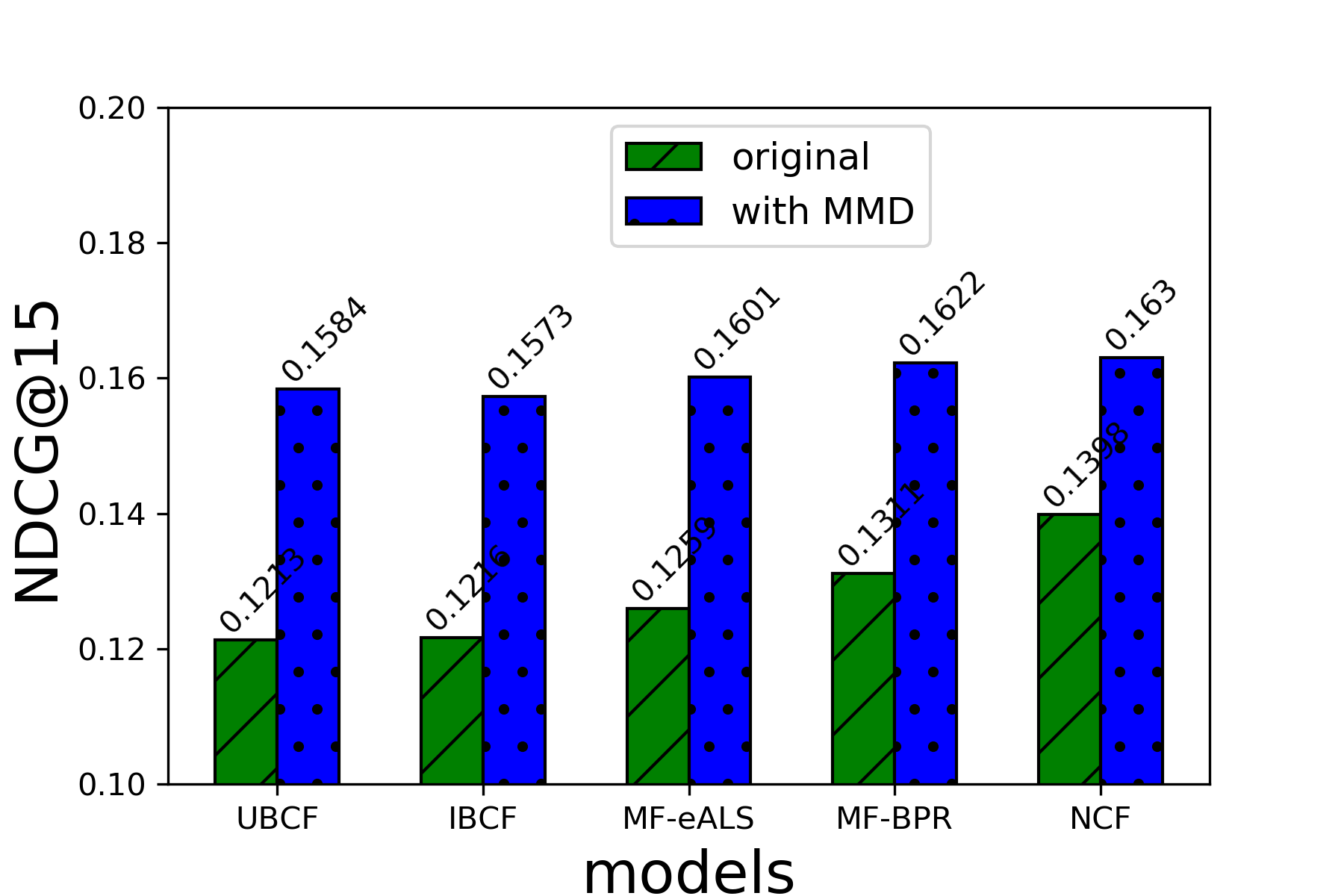} 
			\vspace{-10pt}     
		}
		\subfigure[NDCG@15 on Taobao] { \label{fig:subfig:d7}     
			\includegraphics[width=0.47\columnwidth]{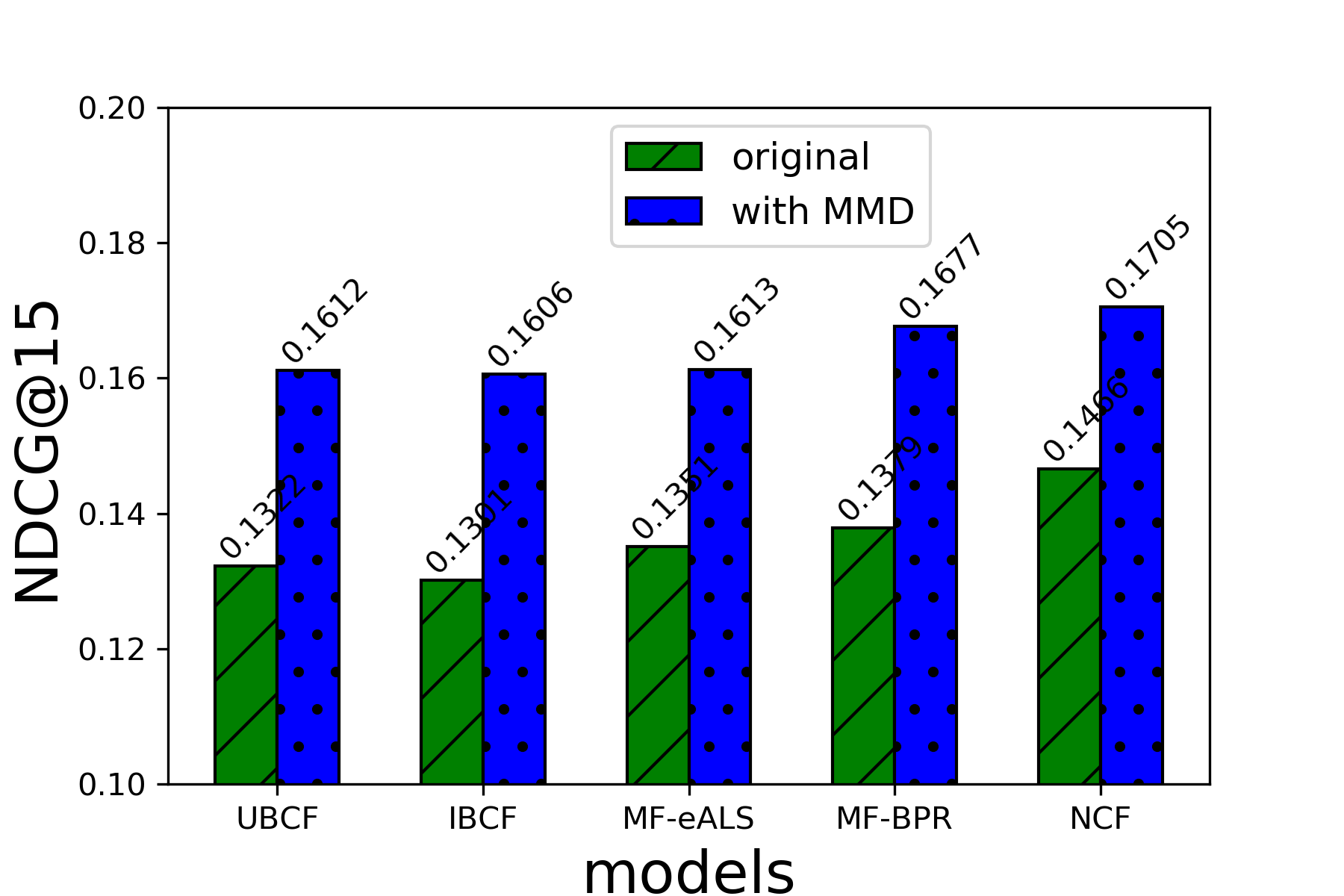} 
			\vspace{-10pt}  
		}     
		\subfigure[NDCG@15 on Jingdong] { \label{fig:subfig:d8}     
			\includegraphics[width=0.47\columnwidth]{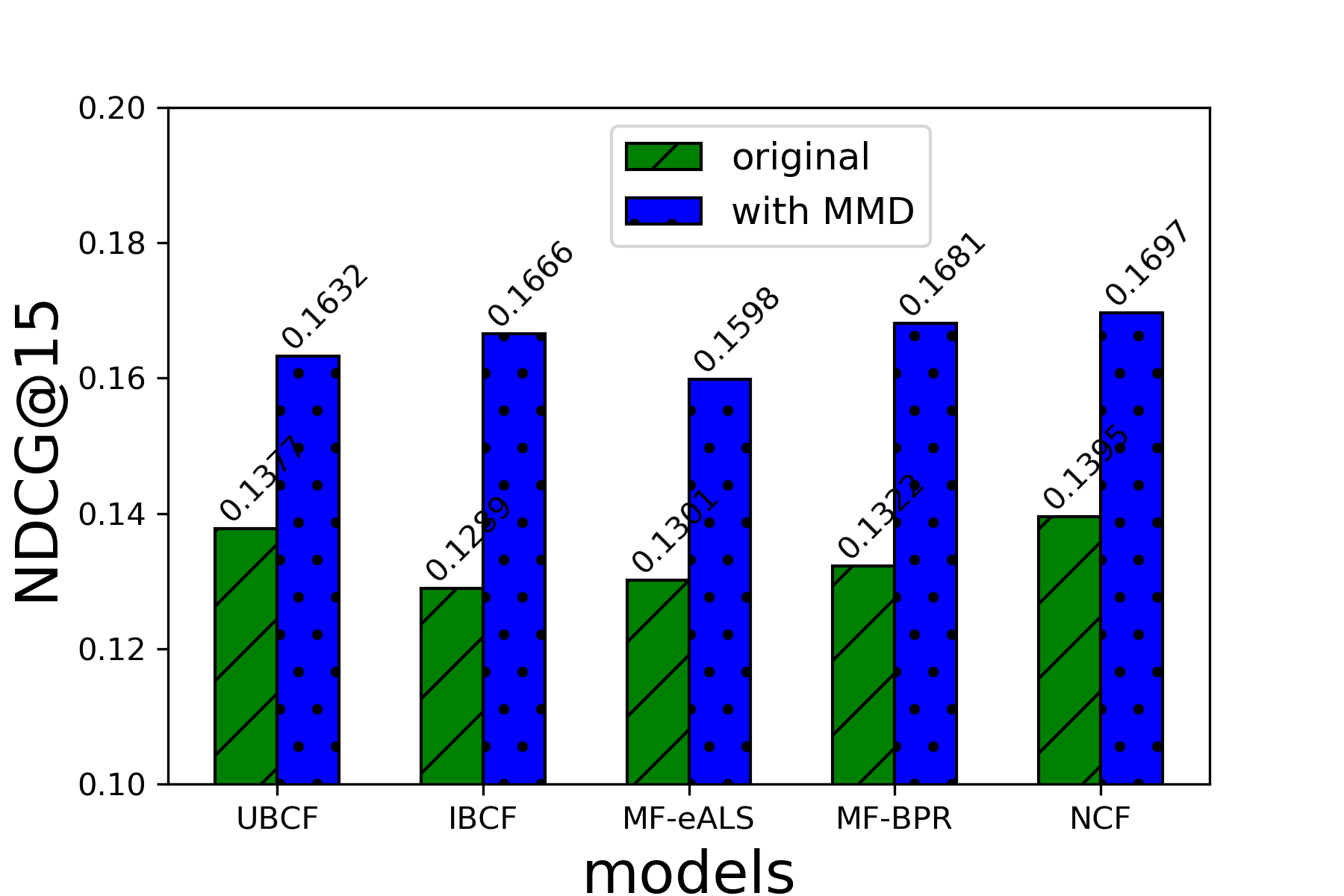} 
			\vspace{-10pt}     
		}   
		\caption{Top-N recommendation performance (NDCG) of different recommendation models with/without MMD as a preprocessing.}     
		\label{MMD7}     
	\end{figure*}
	\subsection{Recommender System Enhancement (RQ4)}
	In this subsection, we validate the effect of MMD on improving recommender systems. Without loss of generality, we conduct our MMD on different recommender system models, such as User-based collaborative model (UBCF) \cite{D36gong2010collaborative}, Item-based collaborative model (IBCF) \cite{D38sarwar2001item}, Matrix Factorization (MF-eALS) \cite{D37koren2009matrix,D39he2016fast}, Bayesian personalized ranking (MF-BPR) \cite{D35rendle2009bpr} and a state-of-the-art neural network-based model: neural collaborative filtering (NCF) \cite{D14he2017neural}. The details of these models can be found in the literature. And we tune the models carefully to achieve their best performance, respectively. To validate the effect of MMD, we fix all the parameters for all the models. The only difference is in terms of the input: the models input the datasets (rating matrices), which drop the 50 professional malicious users detected by MMD (with MMD), while the control models input the original datasets, which randomly drop 50 users (original). The input of RSs is only a rating matrix without reviews, splitting datasets as 60\%, 20\%, 20\% as training, test, and validation. 
	
	Hence, we take MMD as a preprocessing for all the recommender models and compare the results by the metric HR@N and NDCG@N. To be specific, we evaluate the performance of the recommender system concerning N=5, 15. 
	
	Fig.\ref{MMD6} and Fig.\ref{MMD7} show the Top-N performance of different recommendation models with or without MMD as a preprocessing. Among all four datasets, MMD improves the recommendation models significantly in terms of HR and NDCG. Specifically, MMD can enhance the performance of HR by 28.7\% on average and HDCG by 17.3\% on average. By deleting professional malicious users, MMD can improve the quality of datasets. Without these professional malicious users' fake feedbacks, the dataset becomes more intuitive to be understood. Because the feedbacks (reviews and ratings) are closer to the users' real opinions on items, MMD greatly benefits the CF-based models (i.e., UBCF and IBCF). Also, the neural network-based model can be enhanced by MMD to learn a proper latent space for users and items and achieve better performance (NCF).
	\section{Related Works}
	In this section, we briefly introduce related works on malicious user detection and metric learning.
	\subsection{Malicious user detection}
	As we define professional malicious users in recommender systems, malicious user detection is a new problem, which is an issue with little attention yet. However, we can treat this detection issue as a special case of abnormal user detection, and some existing works in this area can inspire us \cite{D56DBLP:conf/ndss/WangJG19,D57DBLP:conf/icdm/WangGF17}. In e-commerce, various abnormal users (spammers, shilling group, and frauds) have greatly damaged the systems, and some abnormal user detection models are proposed to tackle this issue. \cite{45wu2015spammers} proposed a hybrid model to detect the spammers through users' profile and relations. \cite{44guo2017detecting} explored spammer detection in big and sparse data. Shilling attacks harm the recommender system by injecting fake profile information of users and items. They cheat the recommendation model, such as Collaborative Filtering and Matrix Factorization \cite{46tong2018shilling}. \cite{47zhou2015shilling} proposed this attack type and gave a basic supervised solution to tackle it. \cite{46tong2018shilling} proposed a convolutional neural network to solve shilling attacks and improved collaborative filtering. Frauds usually give fake reviews to hurt the profits of electronic retailers. \cite{49weng2018online,48akoglu2013opinion} also explored fraud detection in large-scale dataset and real scenarios. Some researches of abnormal user detection utilize the machine learning model to find fake ratings or reviews \cite{D58DBLP:conf/kdd/XiaLGLCS19,D59DBLP:conf/ccs/YuanMGYLSWL19} and achieve an effective result.
	
	However, different from abnormal users above (spammers, shilling group, and fraud), professional malicious users are smarter and craftier. Shilling attacks inject fake ratings or reviews just before the recommendation process \cite{D54DBLP:conf/www/FangG020,D55DBLP:conf/acsac/FangYGL18}, while for PMUs, all the actions that professional malicious users have taken are well-behaved by the rules of e-commerce websites (called masking strategies). They utilize the bug of abnormal detections, without leaving low ratings and negative feedback at the same time, to avoid detections. Then they can make illegal profits and hurt the electronic retailers. Basically, they are ``normal'' users for the existing abnormal user detection models, which makes the professional malicious user detection a critical issue in the recommender system area.
	\subsection{Metric learning}
	To learn the complex relationships between users' attributes and sentiment gap, we employ the idea of metric learning \cite{D25xing2003distance}. Metric learning is a research spot for image recognition, clustering and recommendation system \cite{18Ye2018Fast,19Li2018Semi,20Wang2018Robust,21Sui2018Convex,22Zuo2017Distance}. The key to metric learning is how to learn a proper set of metrics (such as Euclidean distance or other distance metrics) to represent the relationships between different entities. As a result, some different distances are explored to understand the informational data. In \cite{D50davis2007information}, the authors presented an information-theoretic approach to learn a Mahalanobis distance function for complex data. While \cite{D51globerson2006metric} improved this Mahalanobis distance function for use in classification tasks. To be specific, \cite{D52weinberger2006distance} showed how to learn a Mahanalobis distance metric for k-nearest neighbor (kNN) classification by semidefinite programming. Meanwhile, some research focuses on the constraints for metric learning with large-scale data \cite{D53koestinger2012large}. With the predefined form of the metric matrix, metric learning can achieve a proper distance for the characterization of complex relationships.
	
	Hence, metric learning is usually applied in the computer vision area, in which a deep transfer metric learning method for cross-domain visual recognition was proposed \cite{23Hu2015Deep}. Because of its ability to measure the latent relations between users and items, metric learning is also widely used in recommender systems. CML \cite{1hsieh2017collaborative} directly uses metric learning to embed the relationships between users and items. And IML \cite{6ijcai2018-389} proposes a practical framework to accelerate the learning process. In a word, metric learning has shown great potential to improve the relation representation.
	\section{Conclusion}
	
	In this work, we first defined the professional malicious users (PMUs), who give fake feedbacks to confuse the normal users, hurt the recommender systems, and make illegal profits. We noticed that the traditional outlier detections could not be applied in the recommender system area to detect these professional malicious users because of their professional masking strategies (never give negative reviews and low ratings at the same time). Also, supervised detection models could not work well on PMU detection for the lack of labels. To address the professional malicious user detection issue, we presented a new unsupervised multi-modal learning model named MMD. By utilizing both reviews and ratings simultaneously, MMD obtained a proper metric to cluster users and detected professional malicious users. Extensive results on four real-world datasets demonstrated the effectiveness and strength of our method and the improvement by applying our method for recommender systems.
	
	In essence, MMD is a generic solution, which can not only detect the professional malicious users that are explored in this paper but also serve as a general foundation for malicious user detections. With more data, such as image, video, or sound, the idea of MMD can be instructive to detect the sentiment gap between their title and content, which has a bright future to counter different masking strategies in different applications. Moreover, we will incorporate multimedia data into our model and consider the effect of contexts, such as consuming time, clicks, and other interactions. At last, we are very interested in building an online professional malicious user detection model that utilizes the recent advances in human-machine interactions.
	
	\section*{Acknowledgment}
	This work is supported by NSFC 91746301, and the National Natural Science Foundations of China under Grant No. 61772230 and No. 61972450, Natural Science Foundation of China for Young Scholars No. 61702215 and No. 62002132, China Postdoctoral Science Foundation No. 2017M611322 and No. 2018T110247, and Changchun Science and Technology Development Project No.18DY005.
	\scriptsize
	\bibliographystyle{IEEEtrans}
	\bibliography{yuanbox}
	\vspace{-30pt}
	\begin{IEEEbiography}[{\includegraphics[width=1in,height=1.25in,clip,keepaspectratio]{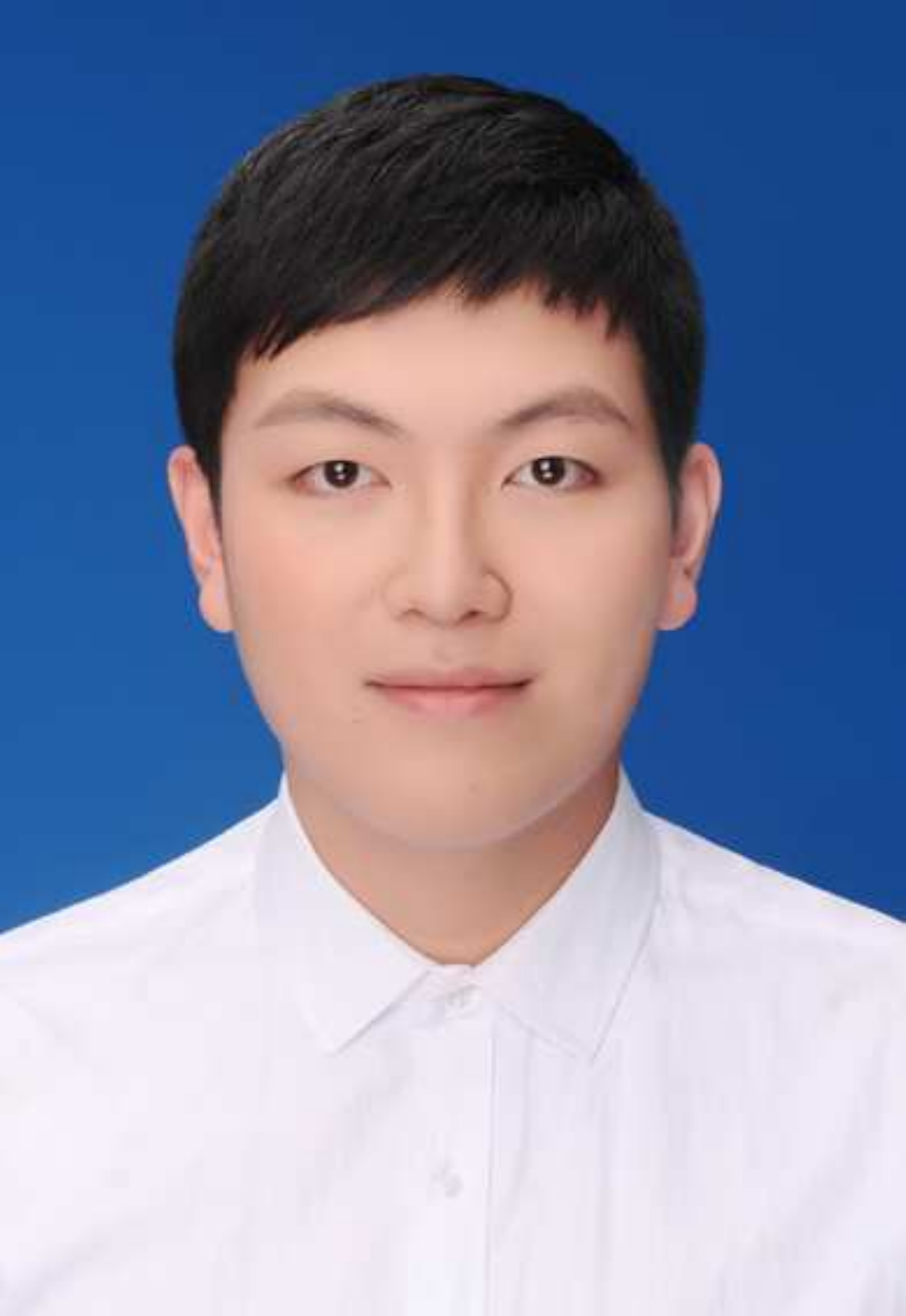}}]
		{Yuanbo Xu} received his B.E. degree in computer science and technology from Jilin University, Changchun, in 2012, his M.E. degree in computer science and technology from Jilin University, Changchun, in 2015, and his Ph.D. in computer science and technology from Jilin
		University, Changchun, in 2019. He is currently a Postdoc in the Department of Artificial Intelligence at Jilin University, Changchun. His research interests include applications of data mining, recommender system, and mobile computing. He has published some research results on journals such as TMM, TNNLS and conference as ICDM.
	\end{IEEEbiography}
	\vspace{-30pt}
	\begin{IEEEbiography}[{\includegraphics[width=1in,height=1.25in,clip,keepaspectratio]{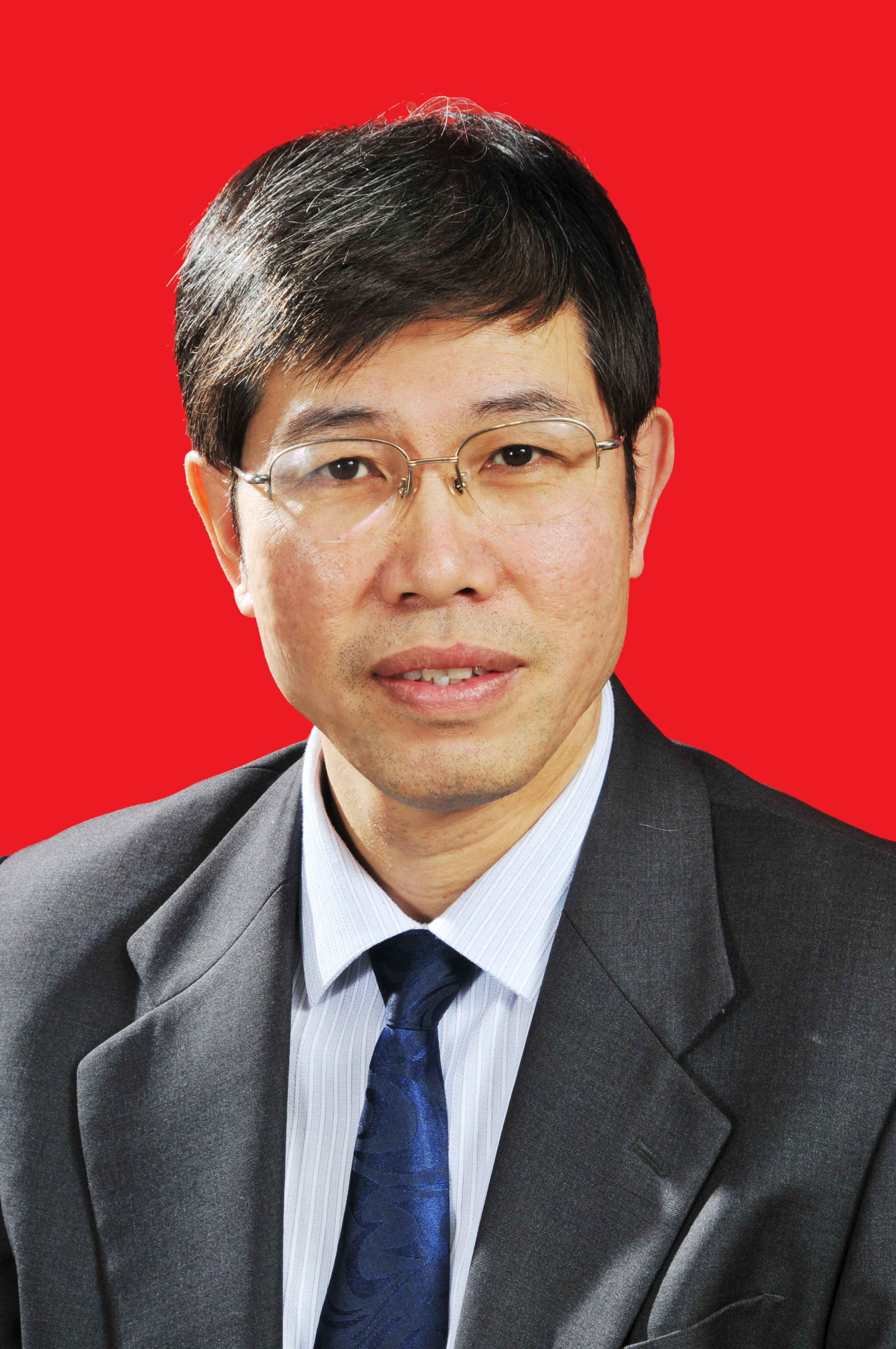}}]
		{Yongjian Yang} received his B.E. degree in automatization from Jilin University of Technology, Changchun, Jilin, China, in 1983; and M.E. degree in Computer Communication from Beijing University of Post and Telecommunications, Beijing, China, in 1991; and his Ph.D. in Software and theory of Computer from Jilin University, Changchun, Jilin, China, in 2005.  He is currently a professor and a PhD supervisor at Jilin University, Director of Key lab under the Ministry of Information Industry, Standing Director of Communication Academy, member of the Computer Science Academy of Jilin Province. His research interests include: Theory and software technology of network intelligence management; Key technology research of wireless mobile communication and services. He participated 3 projects of NSFC, 863 and funded by National Education Ministry for Doctoral Base Foundation. He has authored 12 projects of NSFC, key projects of Ministry of Information Industry, Middle and Young Science and Technology Developing Funds, Jilin provincial programs, ShenZhen, ZhuHai, and Changchun.
	\end{IEEEbiography}
	\vspace{-30pt}
	\begin{IEEEbiography}[{\includegraphics[width=1in,height=1.25in,clip,keepaspectratio]{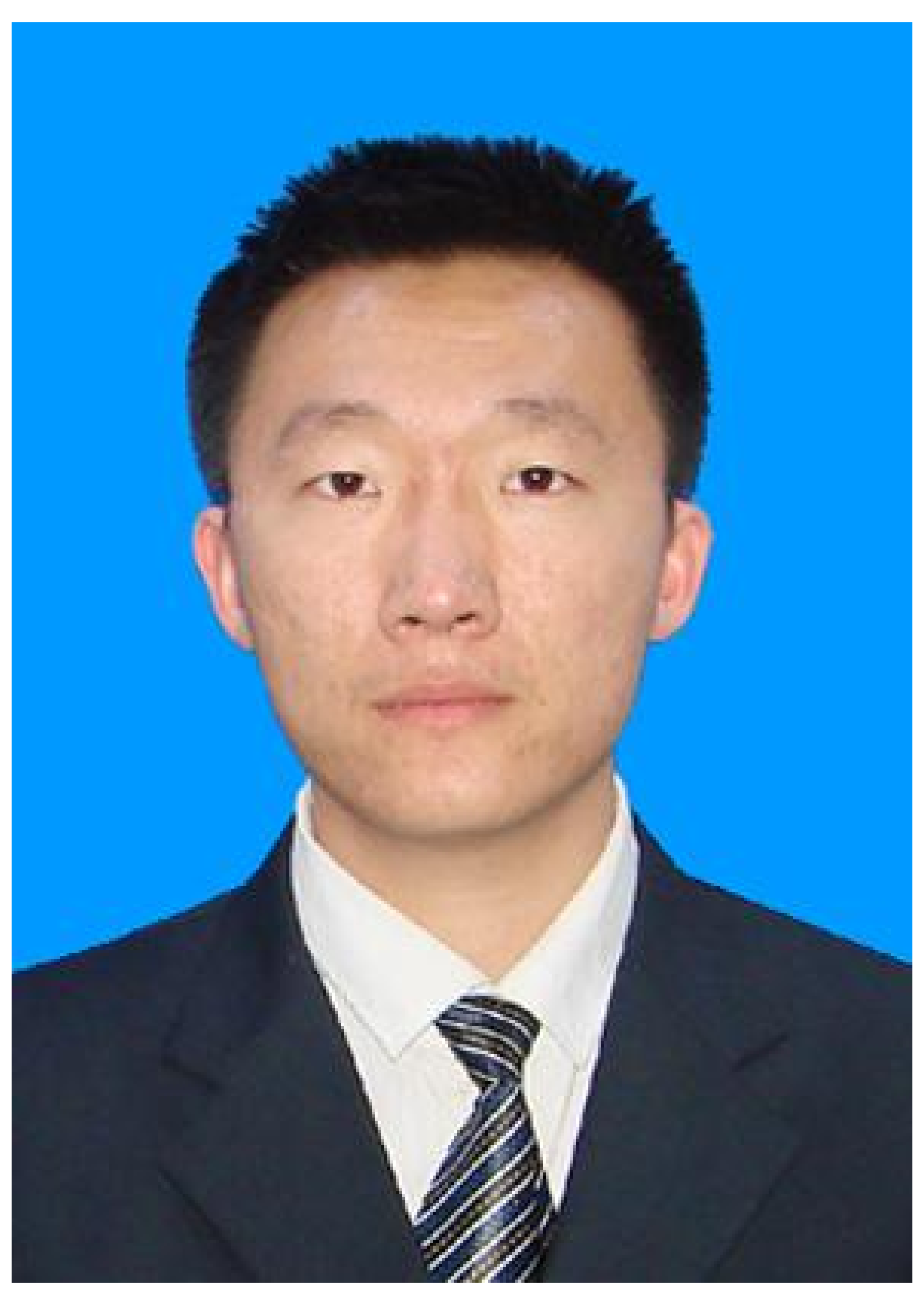}}]
		{En Wang} received his B.E. degree in software engineering from Jilin University, Changchun, in 2011, his M.E. degree in computer science and technology from Jilin University, Changchun, in 2013, and his Ph.D. in computer science and technology from Jilin
		University, Changchun, in 2016. He is currently an Associate Professor in the Department of Computer Science and Technology at Jilin University, Changchun. He is also a visiting scholar in the Department of Computer and Information Sciences at Temple University in Philadelphia. His current research focuses on the efficient utilization of network resources, scheduling and drop strategy in terms of buffer-management, energy-efficient communication between human-carried devices, and mobile crowdsensing.
	\end{IEEEbiography}
	\vspace{-30pt}
	\begin{IEEEbiography}[{\includegraphics[width=1in,height=1.25in,clip,keepaspectratio]{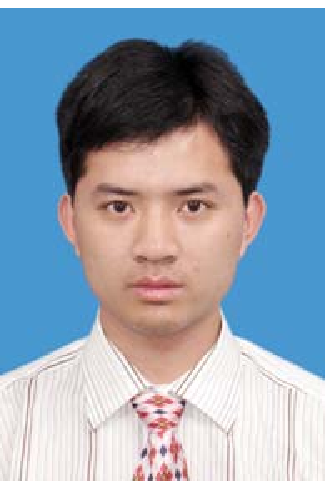}}]
		{Fuzhen Zhuang} is an associate professor in the Institute of Computing Technology, Chinese Academy of Sciences. His research interests include transfer learning, machine learning, data mining, multi-task learning and recommendation systems. He has published more than 100 papers in some prestigious refereed journals and conference proceedings, such as IEEE Transactions on Knowledge and Data Engineering, IEEE Transactions on Cybernetics, IEEE Transactions on Neural Networks and Learning Systems, ACM Transactions on Intelligent Systems and Technology, Information Sciences, Neural Networks, SIGKDD, IJCAI, AAAI, WWW, ICDE, ACM CIKM, ACM WSDM, SIAM SDM and IEEE ICDM. 
	\end{IEEEbiography}
	\vspace{-30pt}
	\begin{IEEEbiography}[{\includegraphics[width=1in,height=1.25in,clip,keepaspectratio]{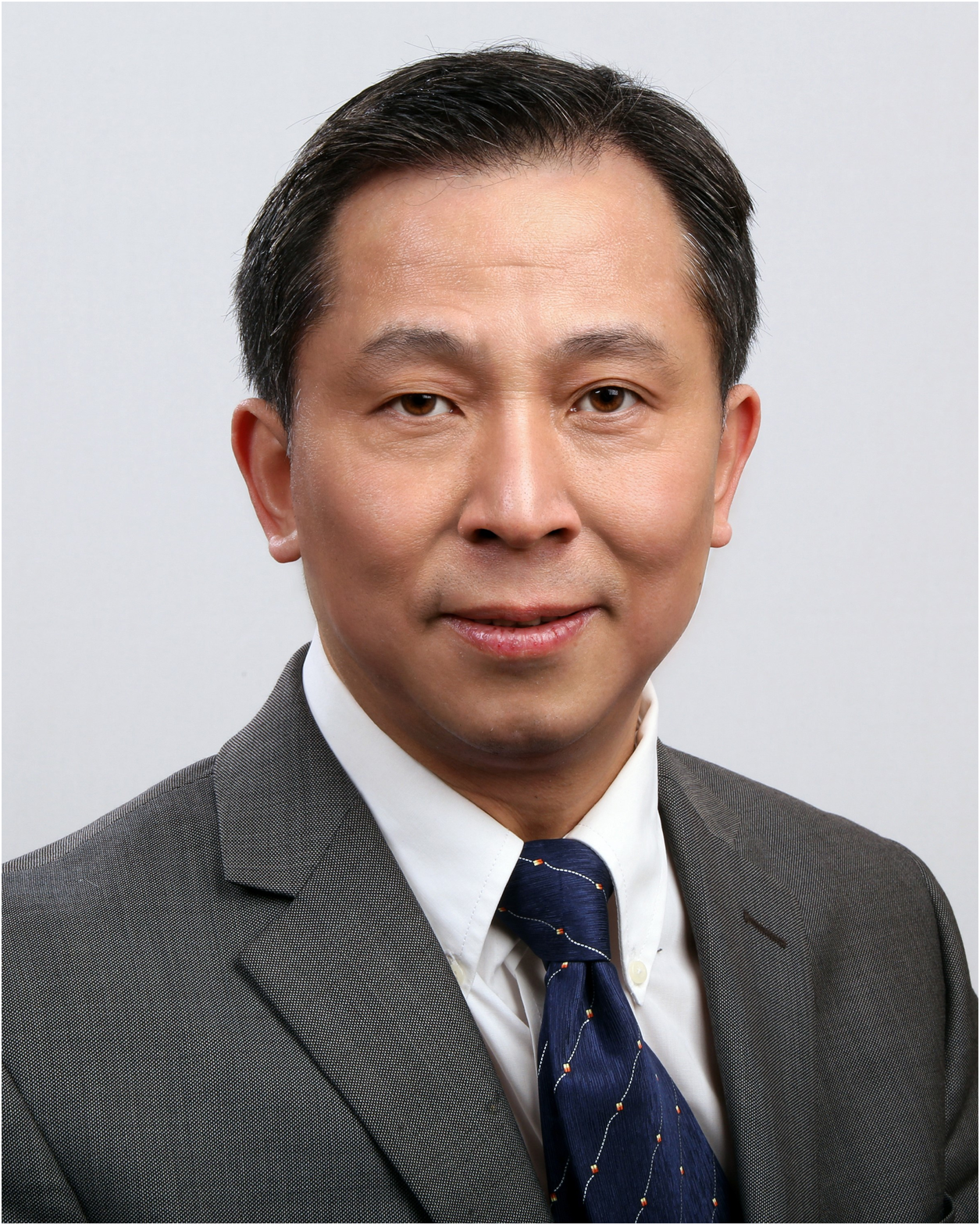}}]  {Hui Xiong} is currently a Full Professor at the Rutgers, the State University of New Jersey, where he received the 2018 Ram Charan Management Practice Award as the Grand Prix winner from the Harvard Business Review, RBS Dean’s Research Professorship (2016), the Rutgers University Board of Trustees Research Fellowship for Scholarly Excellence (2009), the ICDM Best Research Paper Award (2011), and the IEEE ICDM Outstanding Service Award (2017). He received the Ph.D. degree from the University of Minnesota (UMN), USA. He is a co-Editor-in-Chief of Encyclopedia of GIS, an Associate Editor of IEEE Transactions on Big Data (TBD), ACM Transactions on Knowledge Discovery from Data (TKDD), and ACM Transactions on Management Information Systems (TMIS). He has served regularly on the organization and program committees of numerous conferences, including as a Program Co-Chair of the Industrial and Government Track for the 18th ACM SIGKDD International Conference on Knowledge Discovery and Data Mining (KDD), a Program Co-Chair for the IEEE 2013 International Conference on Data Mining (ICDM), a General Co-Chair for the IEEE 2015 International Conference on Data Mining (ICDM), and a Program Co-Chair of the Research Track for the 2018 ACM SIGKDD International Conference on Knowledge Discovery and Data Mining. He is an IEEE Fellow and an ACM Distinguished Scientist.
	\end{IEEEbiography}
\end{document}